\documentclass[preprint,authoryear,12pt]{elsarticle}
\usepackage{amssymb}
\usepackage{color}
\usepackage{amsmath}
\usepackage{subfigure}
\usepackage{graphicx}
\usepackage{graphics}
\usepackage{booktabs}
\usepackage{threeparttable}
\usepackage{appendix}
\usepackage{geometry}
\newcommand{\tabincell}[2]{\begin{tabular}{@{}#1@{}}#2\end{tabular}}
\usepackage[displaymath]{lineno}
\usepackage{setspace}
\usepackage{hyperref}
\usepackage{soul}
\usepackage{lscape}
\usepackage{multirow}
\usepackage{caption}
\DeclareCaptionLabelFormat{cont}{#1~#2\alph{ContinuedFloat}}
\usepackage{mathpazo}

\journal{Transportation Research A, \url{https://doi.org/10.1016/j.tra.2022.11.013}}

\begin{document}

\begin{frontmatter}

\title{Investigating day-to-day route choices based on multi-scenario laboratory experiments. 
Part I: Route-dependent attraction and its modeling}

\author[rvt1,rvt3]{Hang Qi\corref{firstauthor}}

\author[rvt2,rvt3]{Ning Jia\corref{firstauthor}}

\author[rvt4]{Xiaobo Qu}

\author[rvt5]{Zhengbing He\corref{cor1}}

\ead{he.zb@hotmail.com}

\address[rvt1]{Institute for Advanced Studies in Finance and Economics, Hubei University of Economics, China}
\address[rvt2]{Institute of Systems Engineering, College of Management and Economics, Tianjin University, China}
\address[rvt3]{Laboratory of Computation and Analytics of Complex Management Systems (Tianjin University), China}
\address[rvt4]{State Key Laboratory of Automotive Safety and Energy, Tsinghua University, China}
\address[rvt5]{Beijing Key Laboratory of Traffic Engineering, Beijing University of Technology, China}

\cortext[firstauthor]{Co-first author}

\cortext[cor1]{Corresponding author}

\begin{spacing}{1.1}
\begin{abstract} 
In the area of urban transportation networks, a growing number of day-to-day (DTD) traffic dynamic theories have been proposed to describe the network flow evolution, and an increasing amount of laboratory experiments have been conducted to observe travelers' behavior regularities. However, the ``communication" between theorists and experimentalists has not been made well. This paper devotes to 1) detecting unanticipated behavior regularities by conducting a series of laboratory experiments, and 2) improving existing DTD dynamics theories by embedding the observed behavior regularities into a route choice model. First, 312 subjects participated in one of the eight decision-making scenarios and make route choices repeatedly in congestible parallel-route networks. Second, three route-switching behavior patterns that cannot be fully explained by the classic route-choice models are observed. Third, to enrich the explanation power of a discrete route-choice model, behavioral assumptions of route-dependent attractions, i.e., route-dependent inertia and preference, are introduced. An analytical DTD dynamic model is accordingly proposed and proven to steadily converge to a unique equilibrium state. Finally, the proposed DTD model could satisfactorily reproduce the observations in various datasets. The research results can help transportation science theorists to make the best use of  laboratory experimentation and to build network equilibrium or DTD dynamic models with both real behavioral basis and neat mathematical properties.

\end{abstract}

\end{spacing}


\end{frontmatter}



\newpage

\begin{spacing}{1.5}

\section{Introduction}

Traffic equilibrium has been extensively studied due to its significant role in long-term transportation planning. However, real-world traffic in a road network often falls into non-equilibrium states due to various reasons. Day-to-day (DTD) traffic dynamics has attracted much attention, since it focuses on drivers' dynamical shift in route choices and the traffic flow evolution process, rather than merely confining to static equilibrium \citep{Yang2009a,He2010c}.

The most appealing feature of DTD models is flexibility, i.e., it allows to integrate a wide range of behavioral rules into a unified framework \citep{Watling2003}, such as 
forecasting and perception \citep{Watling1999,He2012b,Cantarella2016, Xiao2016}, 
satisfaction-based bounded rationality \citep{Guo2011a,Ye2017}, 
prospect-based decision \citep{Xu2011,Wang2013f}, 
effects of marginal utility \citep{Kumar2015,He2016b}
and social interactions \citep{Wei2016}. 
Although those DTD models greatly enrich our knowledge of traffic evolution and the corresponding equilibrium states, few of them offer empirical or experimental evidence to justify the necessity and validity of these behavioral assumptions.

Empirical research in the real world is challenging, mainly due to the difficulty of the collection of high-quality network-wide traveling and route choice data \citep{Guo2011a,Zhu2011,He2012b,Han2021} and the fact that key variables and scenarios cannot be controlled in real-world data-based research. 
To overcome the difficulty, one of the popular ways is to use the stated choice approach that asks participants to make choices in designed scenarios. 
However, the weakness of the stated choice approach is that participants do not have to take the consequences of their choices, which is likely to result in response biases \citep{hensher2010hypothetical,fifer2014hypothetical}.

Laboratory choice experiments, in which circumstances and contexts can be manipulated with actual consequences, are an enhancement of the stated choice approach. 
They allow a wide range of variables and theories to be examined in controlled scenarios and more detailed data to be collected in verifying theories and models, although they cannot fully reproduce participants' behavior in the real world.
Therefore, laboratory experiments have been widely accepted as a major tool in economic research \citep{Dixit2017,Rapoport2019jebo}.

Compablack to theoretical research, the number of experimental studies on DTD dynamics is limited.
\cite{Iida1992} may be the first experimental research, in which a parallel-route network was employed to test travelers' route choices. 
It found that the mean route choice is close to the equilibrium, while there are considerably large round-to-round fluctuations. 
Likewise, \cite{Selten2007a} reported that the pure equilibrium is approximately achieved based on the experiments in a two-route network, although fluctuations always exist. 
Subjects' choice-making behavior around the equilibrium point was analyzed, which explains why the experimental data keeps fluctuating without a ``smooth" equilibrium. 
\cite{Meneguzzer2013} extended to a three-route network and found a high degree of volatility with no significant relationship between the switching frequency and subjects' characteristics.

Some recent studies attempted to associate experimental results with traditional network equilibrium theory. 
\cite{Dixit2014} attempted to find the best suitable equilibrium theory for describing aggregated route choices in endogenous congestion. The results suggested that the stochastic user equilibrium (SUE) is better than the deterministic user equilibrium (DUE) in terms of the prediction of average choices and the variances in choices.
\cite{Zhao2016c} proposed a satisfaction-based boundedly rational user equilibrium model using the data collected from a two-route experiment. 
Considering the importance of evaluating classic DTD models using laboratory choice experiments, 
\cite{Ye2018} examined several deterministic DTD models based on a virtual DTD route choice experiment, and several theoretical hypotheses were investigated, including the nonlinear effects of path flows, path-dependence, variation of sensitivity, and learning behavior. 
More recently, \cite{meneguzzer2019contrarians} used laboratory experiments to explore different route switching strategies under full and partial information.
The modeling framework in \cite{cantarella1995dynamic} was employed to explain the experimental observations.
Moreover, laboratory choice experiments were also employed to validate route choice models under specific conditions, such as 
the cumulative prospect theory-based route choice model with friends' information sharing \citep{zhang2018cumulative} and
the reinforcement learning-based route choice model in a disrupted network with real-time information \citep{yu2019learning}.

In summary, the existing studies usually use just a single or two scenarios and dozens of rounds in total  \citep{Iida1992,Selten2007a,Zhao2016c,zhang2018cumulative,meneguzzer2019contrarians},
and most of the proposed models belong to static equilibrium models \citep{Dixit2014,Zhao2016c} or not analyzable \citep{Selten2007a,Ye2018,yu2019learning}.

To investigate the DTD route choice behavior more explicitly, we are among the first studies, like  \citet{Wang2021}, who carried out a series of laboratory experiments with various route-choice scenarios.
The rich scenarios and large sample sizes allow us to detect stable behavioral regularities and test the generality of the proposed model with higher reliability.
Specifically, eight scenarios with different number of routes  and cost functions were designed, and each scenario was conducted among three to five groups of subjects who made the same route-choice decisions iteratively 'day by day' for 75 to 110 times.
The main findings and contributions of the paper are as follows.
\vspace{-0.3cm}
\begin{itemize}
\setlength{\itemsep}{0pt}
\setlength{\parsep}{0pt}
\setlength{\parskip}{0pt}

    \item Unanticipated behavioral regularities, namely route-dependent inertia and route-dependent preference, are observed from the individual-level experimental data.
    They are not only inconsistent with classic assumptions of route-choice behavior models (such as perfect rationality or random utility maximization assumption), but also different from the flow-based path-dependent preferences in most recent studies \citep{Cantarella2016,Ye2018}. Therefore, the observations complement the existing route choice theory.

    \item 
    The observed behavioral regularities are integrated into a random utility maximum (RUM) based choice model and a discrete choice-based DTD dynamic model is proposed. 
    The model not only has good theoretical properties, but also can reproduce the experimental observations including the switching rate (i.e., the proportion of travelers switching from one route to another) and the equilibrium flow. 
    Moreover, the model satisfactorily fits the experimental data reported in another existing research, further confirming the external validation of the proposed model.
\end{itemize}

The remainder of the paper is organized as follows. 
Section \ref{sec:Experiments} first introduces the laboratory choice experiments that we conducted.
Section \ref{sec:ExperimentResults} analyzes the experimental data and introduces the newly-observed behavioral patterns. 
To explain the patterns, a discrete choice-based analytical model is proposed in Section \ref{sec:Model}. 
The model is then calibrated and tested in Section \ref{sec:Calibration}. 
Section 6 provides illustrative numerical examples.
Finally, discussions and conclusions are made in Sections \ref{sec:Discussion} and \ref{sec:Conclusion}, respectively.

\section{Day-to-day Route Choice Experiments}\label{sec:Experiments}

We first clarify the following three terminologies: 
\begin{itemize}
\vspace{-3 mm}
    \setlength{\itemsep}{0pt}
    \setlength{\parsep}{0pt}
    \setlength{\parskip}{0pt}

\item
{\bf Scenario}: Each scenario corresponds to a unique experimental design such as network structures and cost functions;

\item
{\bf Session}: In a session, a group of subjects attend an experiment together. 
Each scenario will be examined through more than three different sessions (i.e., with different subjects) to increase the credibility of results;

\item
{\bf Round}: A group of subjects are required to make route-choice decisions repeatedly to simulate the DTD choice process. 
Each repetition is called a ``round", corresponding to ``day" in the real world.

\end{itemize}

\subsection{Subjects}
 
The participants were undergraduate students between 18 and 23 years of age at Tianjin University, China, who volunteered to take part in a decision-making task in return for payoffs contingent on the level of performance. 
A total of 312 subjects (151 males and 161 females) were recruited and 136 of them have driving licenses.
Subjects were randomly assigned into 17 groups with approximately equal proportions of males and females. 
One group was allowed to attend several scenarios but only once for the same scenario.
The subjects were paid at the end of each session, and the reward consisted of two parts: a fixed show-up fee and a bonus contingent on their performance in the experiments.
Eight different scenarios were carried out, and each of them contained multiple repetitions to minimize the influence of random factors.
16 participants were included in each group in Scenario 1-7, and each group contains 24 participants in Scenario 8.
Note that the group sizes of the route-choice experiments are moderately large, which are widely accepted by strategically interactive decision-making experiments \citep{Selten2007a,Rapoport2009,Lindsey2014,Mak2015}.

\subsection{Scenarios}
\begin{figure}[!h]
    \centering
    \subfigure[Symmetric two-route network (Scenario 1)]{
    \includegraphics[width=3.8in]{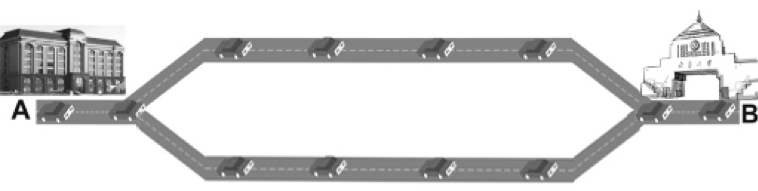}}
    \subfigure[Asymmetric two-route network (Scenarios 2-5)]{
    \includegraphics[width=4in]{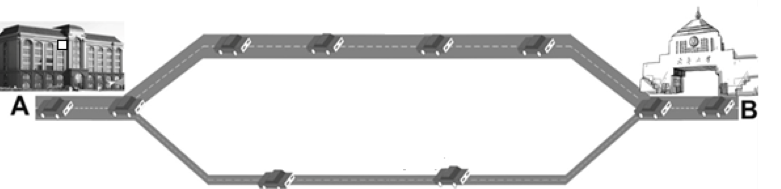}}
    \subfigure[Asymmetric three-route network (Scenarios 6-8)]{
    \includegraphics[width=3.9in]{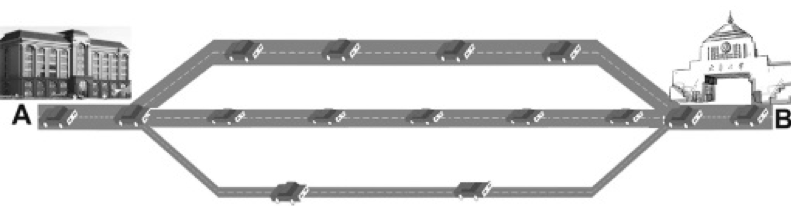}}
    \caption{Three networks that are employed in the eight scenarios.}
    \label{fig:Scenario}
\end{figure}

A network with a single origin-destination (OD) pair was employed (Figure \ref{fig:Scenario}), which is widely adopted by the existing research \citep{Iida1992,Selten2007a}. 
In the scenarios, a fixed number of commuters that were supposed to live in the same community (the origin) must choose one among several possible routes in order to reach a common workplace (the destination) every morning. 
Each route was designed to be susceptible to congestion. 
The participants were instructed to ``arrive at the workplace as quickly as possible". 
The travel time of a given route was assumed to be an increasing function of the route flow.

The eight scenarios (Figure \ref{fig:Scenario} and Table \ref{tab:Parameters}) are designed as follows.
\begin{itemize}
\vspace{-3 mm}
    \setlength{\itemsep}{0pt}
    \setlength{\parsep}{0pt}
    \setlength{\parskip}{0pt}

\item 
{\bf Scenario 1}: Scenario 1 is a baseline with a symmetric two-route network. 

\item 
{\bf Scenarios 2-5}: Scenarios 2-5 extend Scenario 1 to asymmetric two-route networks with different cost functions so as to investigate subjects’ route choice behaviors under different cost feedback. The cost functions are designed referring to the pioneering research in \cite{Selten2007a};

\item 
{\bf Scenarios 6-7}: Since the binary-choice behavior is relatively simple, we carried out Scenarios 6-7 with asymmetric three-route networks, in which three alternative routes were presented to subjects and we expected to observe more behavioral patterns.

\item 
{\bf Scenario 8}: Scenario 8 further extends the configuration to non-linear cost functions and different group sizes, which would demonstrate the robustness of the proposed theoretical model.

\end{itemize}

Table \ref{tab:Parameters} presents the details of each scenario with its unique DUE flow assignment. 
In Scenarios 1-7, the linear cost function was used due to the following considerations.
First, in spite of simplicity, linear cost functions capture the most important aspects of the relationship between travel time and traffic flow. 
Therefore, linear cost functions are used in many existing experiment-based studies, e.g., \cite{Selten2007a},\cite{Zhao2016c}, \cite{Knorr2014} and \citet{Wang2021}. 
Second, the participants did not know the exact travel time function and the linear relation between travel time and demand. 
Third, linear cost functions also help to minimize the factors that may impact on decision-making, which is important to locate the most important driving forces behind route choice behavior. Linear cost functions could capture the most important aspects of the relationship between travel time and traffic flow loaded on routes and it will not induce additional non-linear effects that may cost participants more mental efforts in decision-making.
We further check the robustness of the proposed theoretical model by using Scenario 8 with the classic non-linear Bureau of Public Roads (BPR) cost function and a different group size (i.e., 24 participants).

\begin{table}[!htbp]\footnotesize\center
\setlength{\tabcolsep}{2.5mm}{
\caption{Parameters setting in all scenarios with corresponding DUE solutions.}\label{tab:Parameters}
\begin{tabular}{cccccc}
\toprule
{\footnotesize Scenario} & {\footnotesize Number of routes} & {\footnotesize Cost function$^{*}$} & {\footnotesize DUE$^{**}$} & {\footnotesize Session} & {\footnotesize Number of rounds} \\
\midrule
 1 & 2 & \tabincell{c}{ $c_1= 6 + 2f_1$ \\ $c_2= 6+ 2f_2$ }  &  8\ ,\ 8  & \tabincell{c}{1 \\ 2 \\ 3 }  &  \tabincell{c}{93 \\ 83 \\ 94 }\vspace{2mm}\\
 2 & 2 & \tabincell{c}{ $c_1= 10+ 4f_1$ \\ $c_2= 24+ 6f_2$ }  &  11\ ,\ 5  & \tabincell{c}{1 \\ 2 \\ 3 }  &  \tabincell{c}{90 \\ 90 \\ 98 }\vspace{2mm}\\
 3 & 2 & \tabincell{c}{ $c_1= 5+ 2f_1$ \\ $c_2= 12+ 3f_2$ }  &  11\ ,\ 5  & \tabincell{c}{1 \\ 2 \\ 3 }  &  \tabincell{c}{87 \\ 93 \\ 83 }\vspace{2mm}\\
 4 & 2 & \tabincell{c}{ $c_1= 12+ 4f_1$ \\ $c_2= 24+ 6f_2$ }  &  10.8\ ,\ 5.2  & \tabincell{c}{1 \\ 2 \\ 3 \\ 4 }  &  \tabincell{c}{87 \\ 50 \\ 99 \\78 }\vspace{2mm}\\
 5 & 2 & \tabincell{c}{ $c_1= 6+ 2f_1$ \\ $c_2= 12+ 3f_2$ }  &  10.8\ ,\ 5.2  & \tabincell{c}{1 \\ 2 \\ 3 \\ 4 }  &  \tabincell{c}{89 \\ 84 \\ 95 \\90 }\vspace{2mm}\\
 6 & 3 & \tabincell{c}{ $c_1= 22+ 4f_1$ \\ $c_2= 24+ 6f_2$ \\ $c_3= 30+ 8f_3$}  &  8\ ,\ 5\ , \ 3  & \tabincell{c}{1 \\ 2 \\ 3 \\ 4 \\ 5}  &  \tabincell{c}{85 \\ 83 \\ 104 \\ 97 \\85 }\vspace{2mm}\\
 7 & 3 & \tabincell{c}{ $c_1= 11+ 2f_1$ \\ $c_2= 12+ 3f_2$ \\ $c_3= 15+ 4f_3$}  &  8\ ,\ 5\ , \ 3  & \tabincell{c}{1 \\ 2 \\ 3 \\ 4 \\ 5}  &  \tabincell{c}{87 \\ 84 \\ 105 \\ 91 \\91 }\vspace{2mm}\\
 8 & 3 & \tabincell{c}{ $c_{1}=43.75(1+0.15({f_{1}}/{6})^{2})$ \\ $c_{2}=43.75(1+0.15({f_{2}}/{4})^{2})$ \\ $c_{3}=43.75(1+0.15({f_{3}}/{2})^{2})$}  &  12\ ,\ 8\ , \ 4  & \tabincell{c}{1 \\ 2 \\ 3 \\ 4 \\ 5}  &  \tabincell{c}{83 \\ 80 \\ 83 \\ 88 \\86 }\\

\bottomrule
\end{tabular}}
\begin{tablenotes}\scriptsize
\item[1]$^{*}$ $c_i$ and $f_i$ are the cost and flow of route $i$, respectively.
\item[1]$^{**}$ DUE flow assignment; The unit is ``traveler".
 \end{tablenotes}
\end{table}


\subsection{Procedure}
The experiments were conducted in a computerized laboratory with multiple terminals located in separate cubicles.
After entering the laboratory, each subject drew a poker chip to determine his/her seating. 
Experiment instructions were assigned to subjects at the beginning of each session and the subjects read them at their own places. 
The instructions described the experiments in detail, including the network and the repeated decision-making scenarios, the procedure of the experiment session, the computation of their cash reward dependent on their performance in the game, and the basic rules of the experiment (e.g., the no-talking rule). 
The participants were notified that they are playing a congestion game, in which their rewards depend on others' choices. However, they were NOT informed the exact form of cost functions. The setting completely follows the routines of congestion game experiments in \cite{Selten2007a},\cite{Rapoport2009,Rapoport2014}, etc.
Scenarios 1-7 were carried out during January 2015 and April 2016, and Scenario 8 in October 2019.


The DTD route choice procedure was simulated through a client/server computer program, i.e., a human-in-the-loop system. 
Subjects were required to choose one route from two or three alternatives in each round. 
The travel times of all routes were calculated by the server after all choices were made. 
Then, selected feedback information was presented to subjects, including the travel time of subject's chosen and non-chosen routes in the last round, the payoff in the last round, the cumulative payoff in RMB\footnote{RMB is the unit of Chinese currency, and 7 RMB is approximately equal to 1 US dollar.}, and the historical travel times of all routes. 
Afterwards, the next round began and all subjects were required to make another choice.
Note that disclosing travel time information is to be consistent with the classic DTD model settings that are compared in the research.
In addition, with the deployment of Advanced Traveler Information Systems or navigation apps, such travel time of most alternative routes can be easily accessed in the real world nowadays.

The termination of a session was randomly determined by the server after 75 to 110 rounds, since too many rounds might make subjects exhausted and thus lower the experiment quality. 
This setting is similar to that in the existing experiment-based studies; 
e.g., 200 rounds in \cite{Selten2007a}, 100 rounds in \cite{CHMURA2012}, and 40 rounds in \cite{Rapoport2009}. Both network-level data (i.e., flow and travel time) and individual-level data (i.e., every single subject's decision) were recorded automatically.

At the end of each session, the participants were paid. The reward consisted of two parts: a fixed show-up fee of 15 RMB and a bonus contingent to their performance in the experiments. 
%
The payoff point for each round was equal to the subtraction 
of the travel time in the round
from a fixed reward.
The total payoff points for all rounds were converted to a cash reward with the rate of ``50 points = 1 RMB". On average, the participants earned 98 RMB, with a maximum of 130 RMB and a minimum of 85 RMB.

\subsection{Descriptive Statistics of Experiment Results}


First, for each scenario, the observed mean flows from the same route of all sessions are compared and no significant differences are found through the Wilcoxon sign-ranked test (all $p>0.05$). Thus, the data collected from different sessions of a scenario are pooled together for analysis as done in previous studies such as  \cite{Selten2007a} and \cite{Rapoport2014}.

Second, the descriptive statistics of the experimental data are presented in Tables \ref{tab:Statistics_Time} and \ref{tab:Statistics_Flow}. 
Generally speaking, the means of flow and travel time are all close to the DUE equilibrium points. 
Regarding the variance, it is found in the two-route scenarios (Scenarios 1-5) that the flow variances of the two routes are very close in Table \ref{tab:Statistics_Flow}, turning out that the route with higher cost sensitivity to flow has larger travel time variance. 
In the three-route scenarios (Scenarios 6-8), route 1 has the largest flow variance but the smallest travel time variance, and route 3 has the smallest flow variance but the largest travel time variance, mainly due to different cost sensitivity.

\begin{table}[!htbp]\footnotesize\center
\setlength{\tabcolsep}{3.1mm}{
\caption{Descriptive statistics of travel time on different routes in each scenario.}\label{tab:Statistics_Time}
\begin{tabular}{cccccc cccccc}
\toprule
Scenario & \multicolumn{3}{c}{route 1} & & \multicolumn{3}{c}{route 2} & & \multicolumn{3}{c}{route 3}  \\
\cline{2-4}  \cline{6-8}  \cline{10-12}  
 & DUE & Mean & S.D. & DUE & Mean & S.D. & & DUE & Mean & S.D.  \\
\midrule
 1 & 22 & 22.08 & 3.55 & & 22 & 21.92 & 3.55  & & --- & --- & --- \\
 2 & 54 & 53.59 & 6.09 & & 54 & 54.61 & 9.14  & & --- & --- & --- \\
 3 & 27 & 26.83 & 3.42 & & 27 & 27.25 & 5.13  & & --- & --- & --- \\
 4 & 55.2 & 54.35 & 7.52 & & 55.2 & 56.48 & 11.28 & & --- & --- & --- \\
 5 & 27.6 & 27.21 & 3.90 & & 27.6 & 28.18 & 5.84  & & --- & --- & --- \\
 6 & 54   & 53.86 & 5.99 & & 54   & 54.45 & 8.61 & & 54 & 53.68 & 10.32 \\ 
 7 & 27   & 26.96 & 3.34 & & 27   & 26.47 & 4.64 & & 27 & 27.80 & 5.90  \\
 8 & 70   & 71.11 & 10.11 & & 70  & 70.58 & 14.23 & & 70 & 78.32 & 22.93  \\
\bottomrule
\end{tabular}}
\end{table}

\begin{table}[!htbp]\footnotesize\center
\setlength{\tabcolsep}{3.1mm}{
\caption{Descriptive statistics of route flow on all routes in each scenario.}\label{tab:Statistics_Flow}
\begin{tabular}{cccccc cccccc}
\toprule
Scenario & \multicolumn{3}{c}{route 1} & & \multicolumn{3}{c}{route 2} & & \multicolumn{3}{c}{route 3}  \\
\cline{2-4}  \cline{6-8}  \cline{10-12}  
 & DUE & Mean & S.D. & & DUE & Mean & S.D. & & DUE & Mean & S.D.  \\
\midrule
 1 & 8 & 8.04     & 1.77 & &  8 & 7.96 & 1.77 & & --- & --- & --- \\
 2 & 11 & 10.90   & 1.52 & &  5 & 5.10 & 1.52 & & --- & --- & --- \\
 3 & 11 & 10.92   & 1.71 & &  5 & 5.08 & 1.71 & & --- & --- & --- \\
 4 & 10.8 & 10.59 & 1.88 & &  5.2 & 5.41 & 1.88 & & --- & --- & ---  \\
 5 & 10.8 & 10.61 & 1.95 & &  5.2 & 5.39 & 1.95 & & --- & --- & --- \\
 6 & 8    & 7.96  & 1.50 & &  5   & 5.08 & 1.43 & & 3  & 2.96  & 1.29 \\ 
 7 & 8    & 7.98  & 1.67 & &  5   & 4.82 & 1.55 & & 3  & 3.20  & 1.48 \\
 8 & 12    & 12.1  & 2.26 & &  8   & 7.87 & 2.19 & & 4  & 4.13  & 1.75 \\

\bottomrule
\end{tabular}}
\end{table}

\begin{figure}[!htbp]
    \centering
    \subfigure[Scenario 1]{
    \includegraphics[width=2.8in]{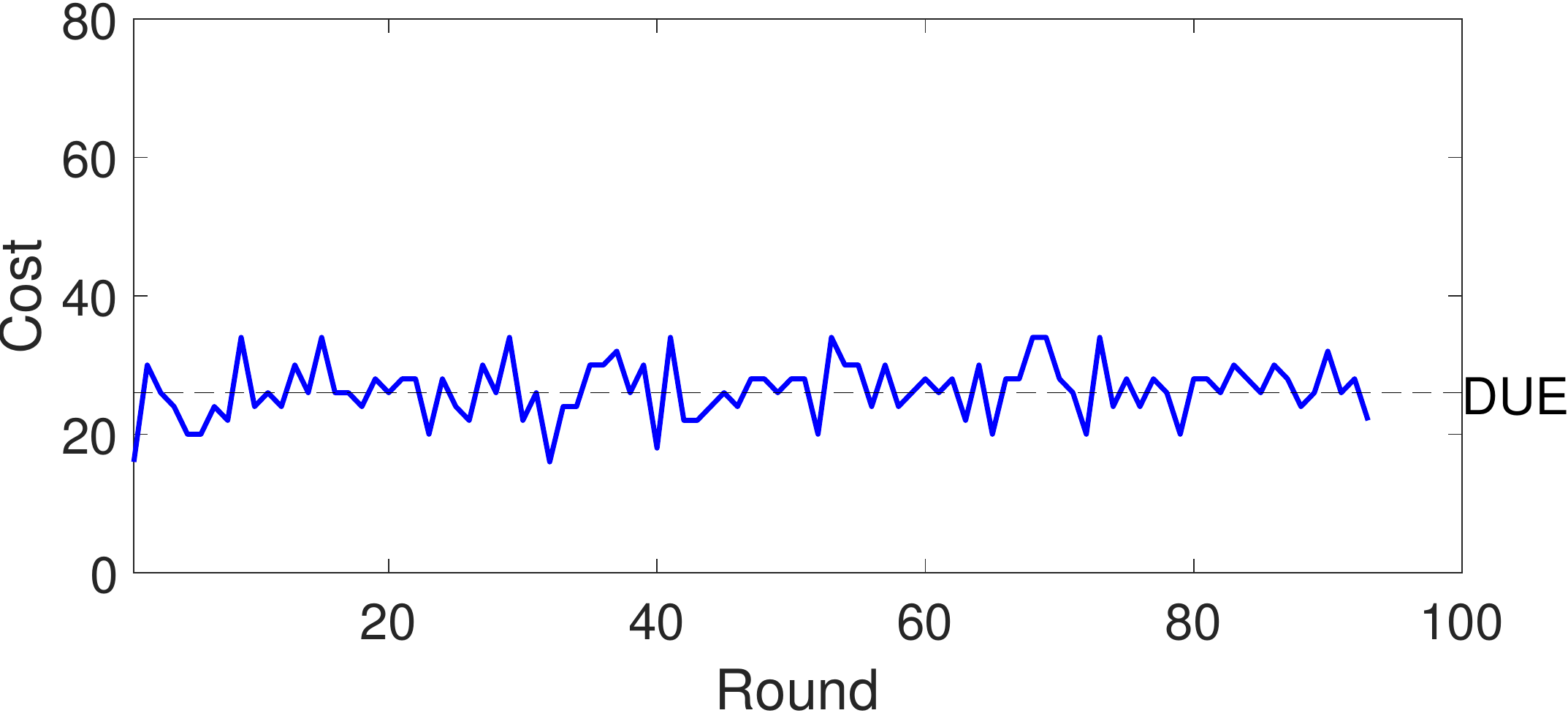}}
    \hspace{5mm}
    \subfigure[Scenario 2]{
    \includegraphics[width=2.8in]{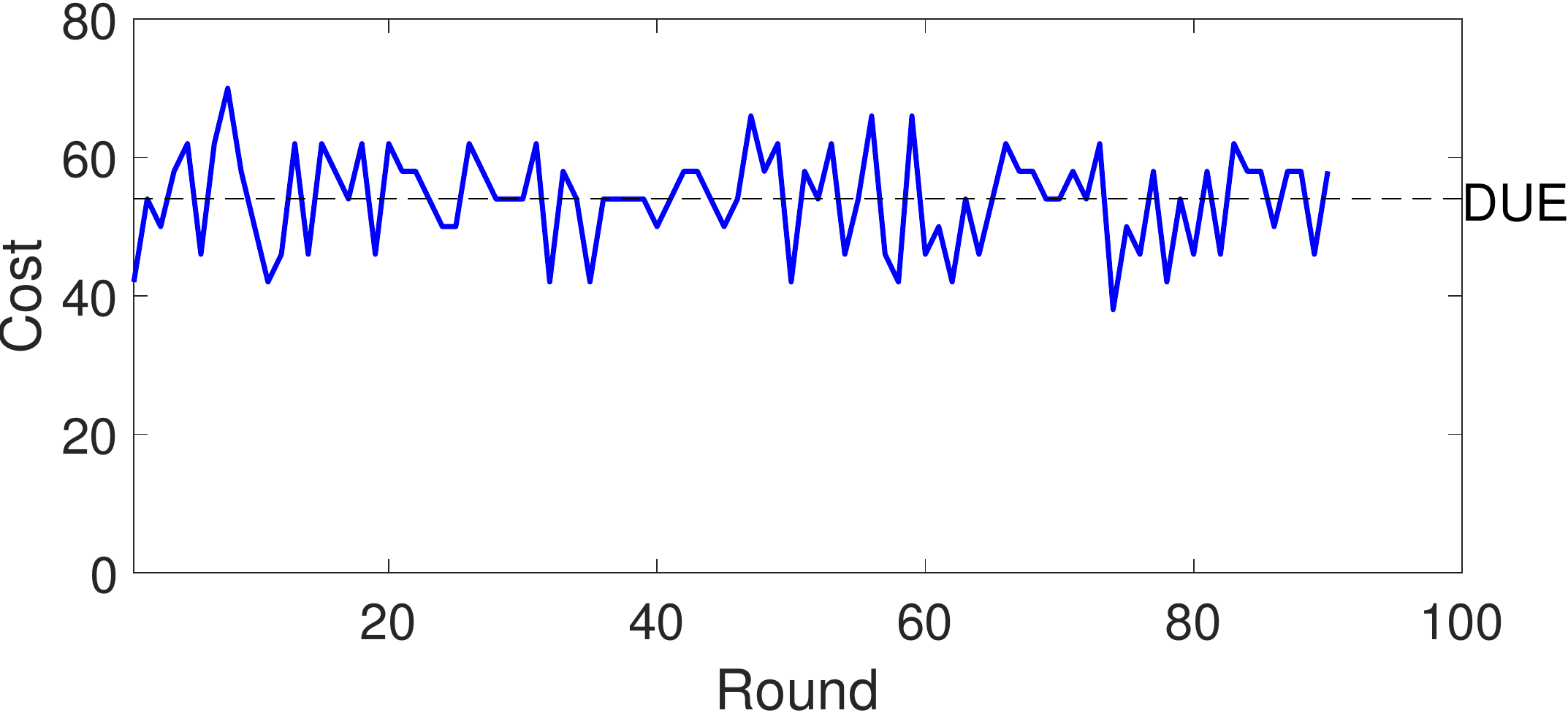}}
    \subfigure[Scenario 3]{
    \includegraphics[width=2.8in]{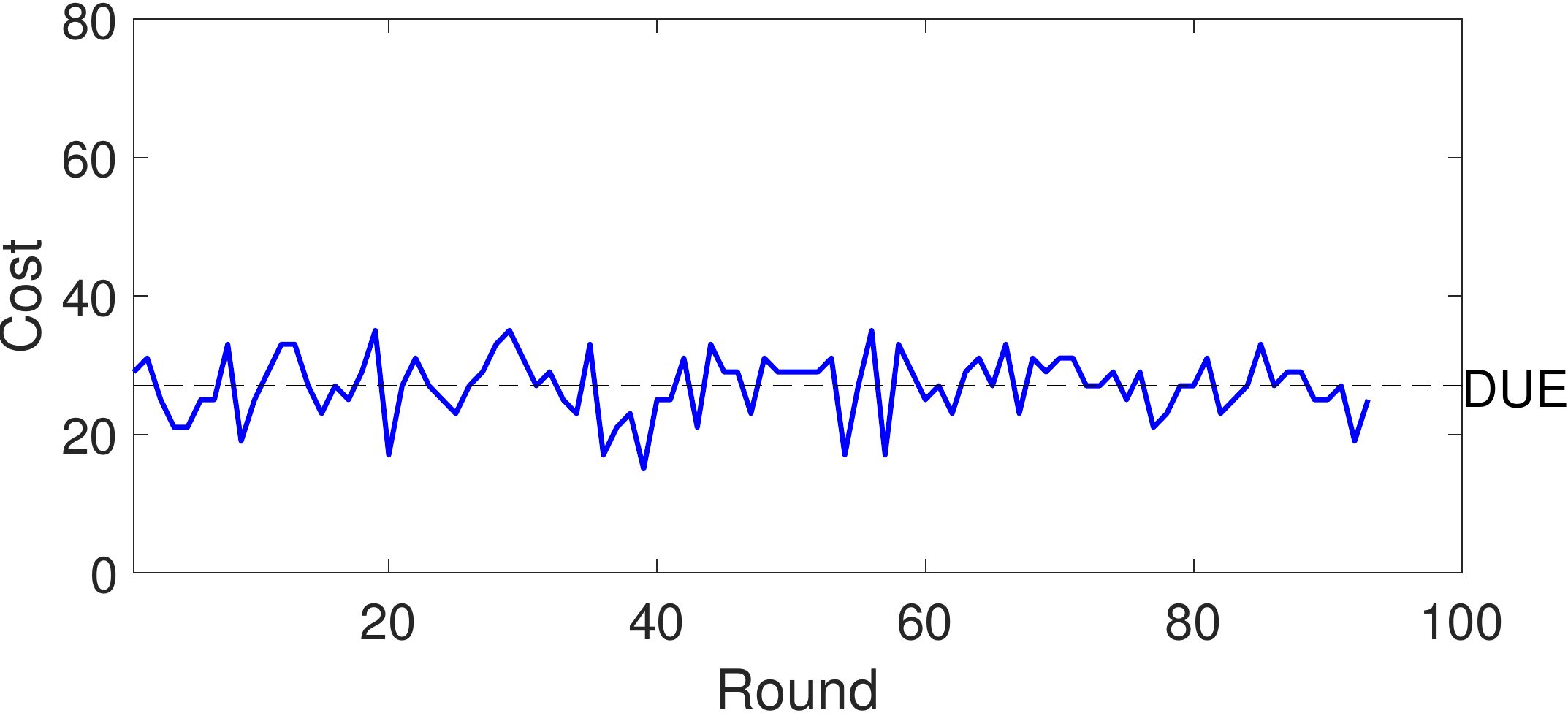}}
    \hspace{5mm}
    \subfigure[Scenario 4]{
    \includegraphics[width=2.8in]{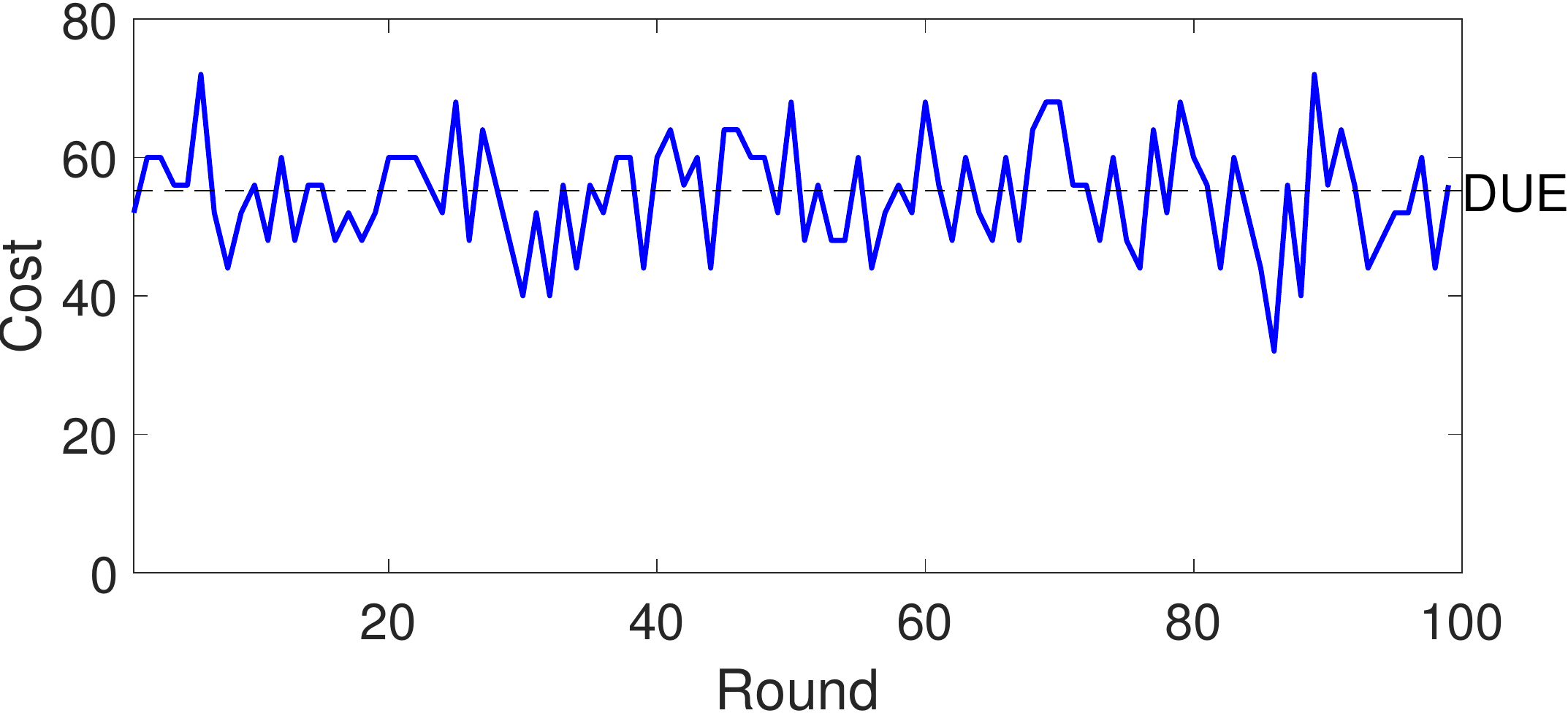}}
    \subfigure[Scenario 5]{
    \includegraphics[width=2.8in]{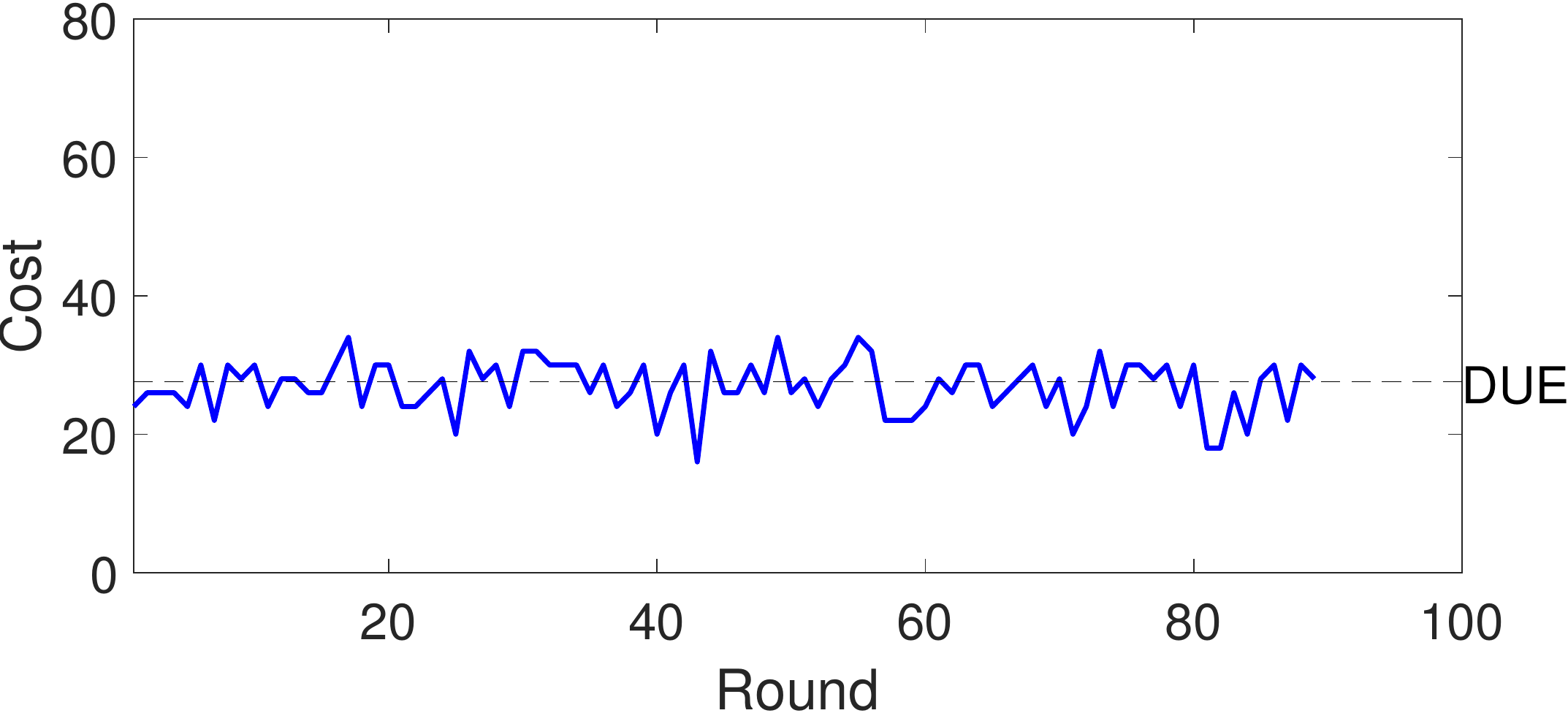}}    
    \hspace{5mm}
    \subfigure[Scenario 6]{
    \includegraphics[width=2.8in]{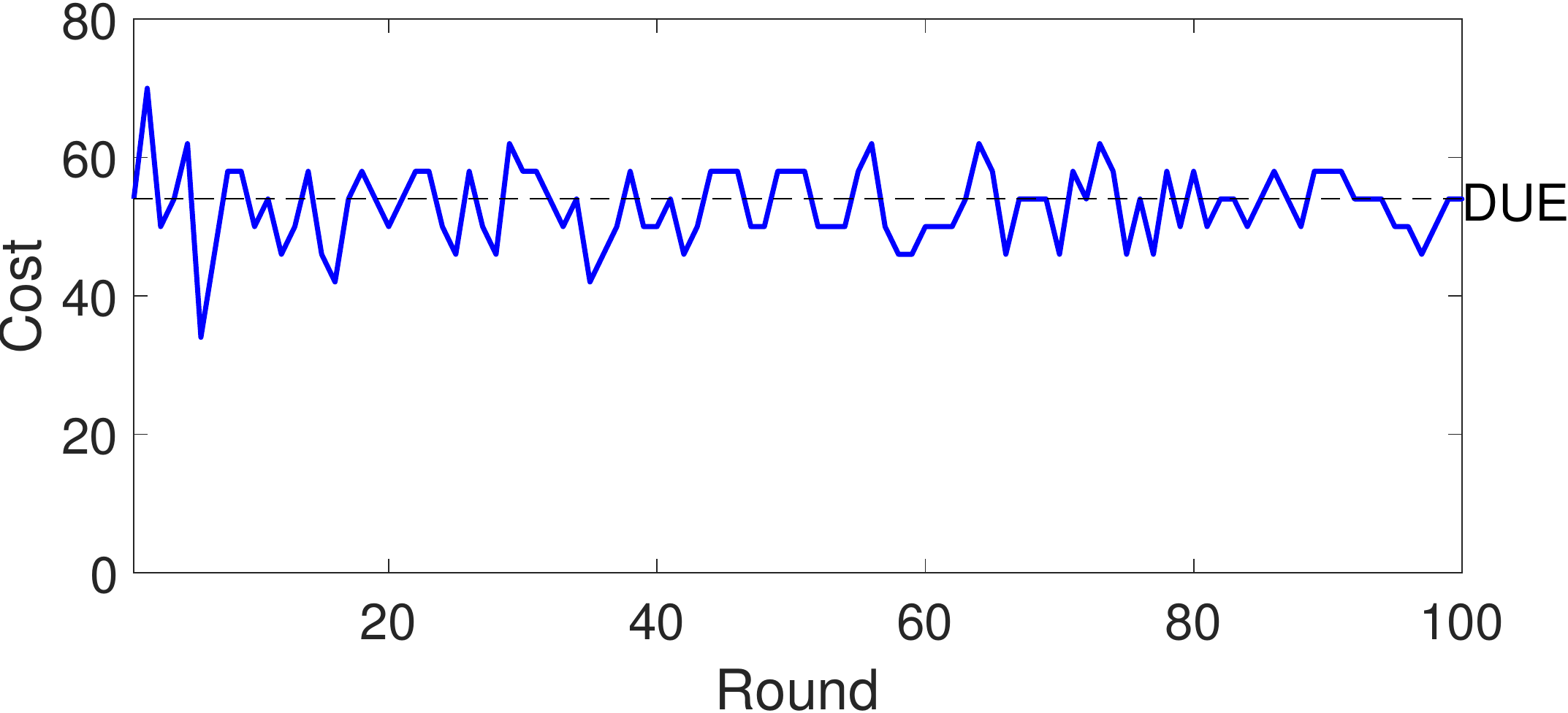}}
    \subfigure[Scenario 7]{
    \includegraphics[width=2.8in]{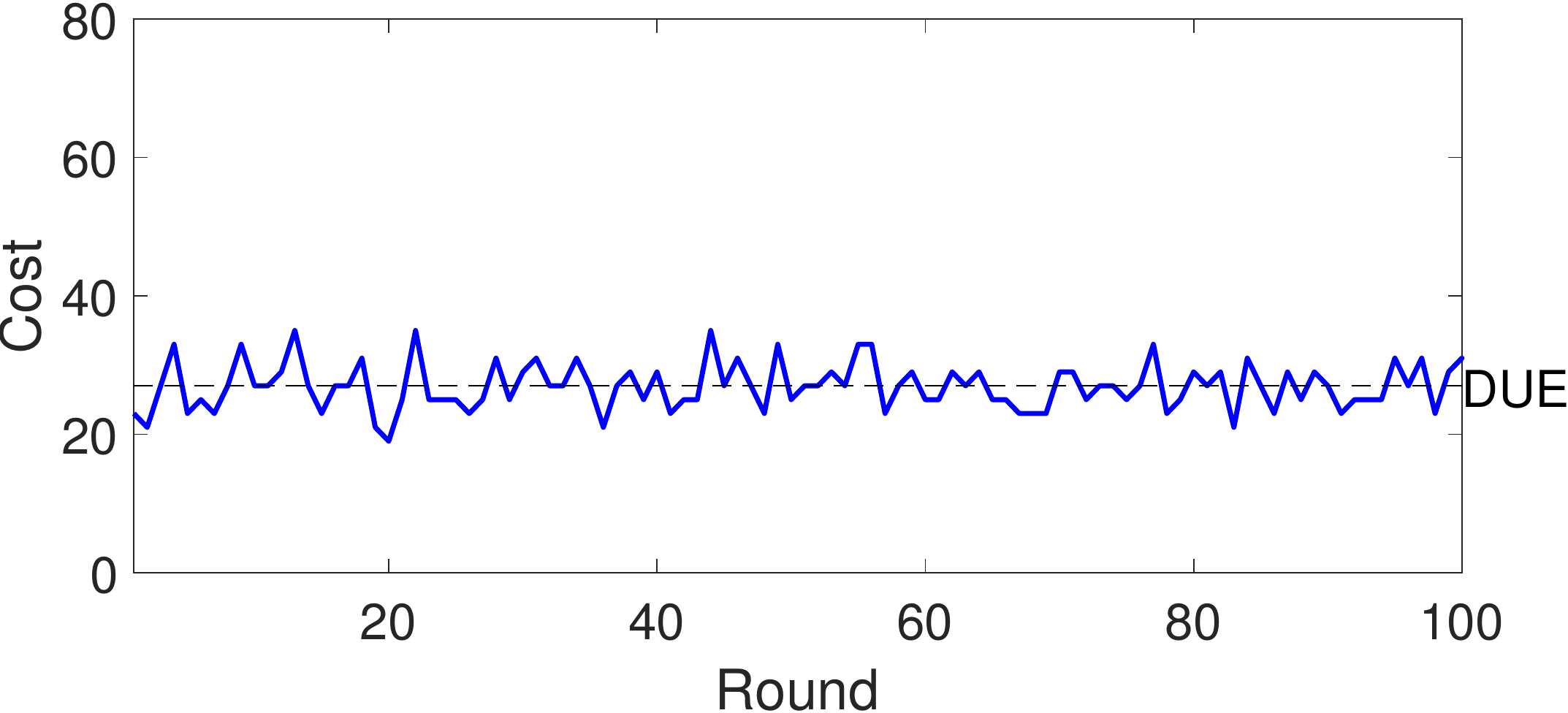}}    
    \subfigure[Scenario 8]{
    \includegraphics[width=2.8in]{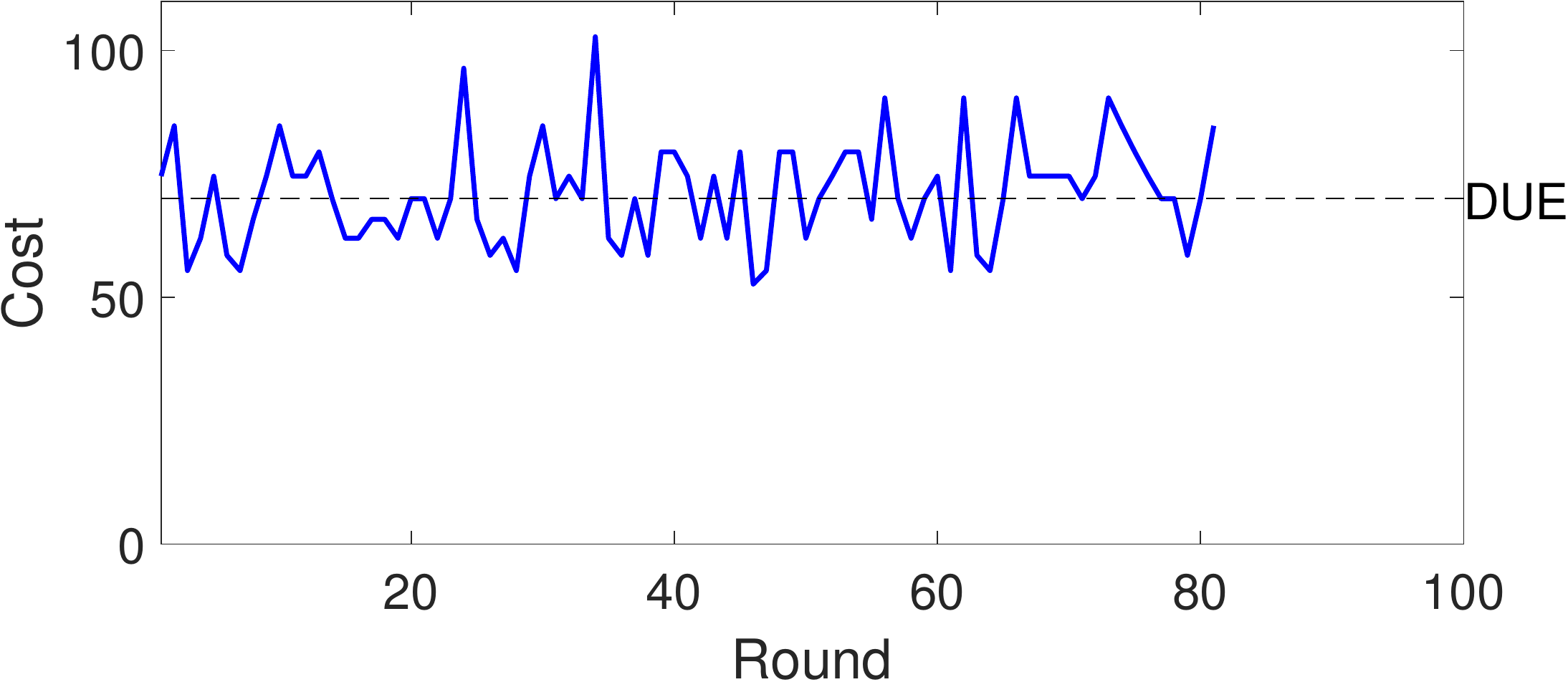}}    
    \caption{Travel time evolution in all scenarios in the experiment. 
    The route cost (i.e., travel time) on the route 1 is shown here as an example.}
    \label{fig:Evolution}
\end{figure}

While the average trends of flow and travel time are close to the DUE equilibrium points, the detailed value in each round keeps fluctuating around the equilibrium point (see Figure \ref{fig:Evolution} for an illustration), as previously reported \citep{Selten2007a,Meneguzzer2013,Dixit2014}.
This research focuses on analyzing and modeling the average trend network flow dynamics, rather than both trend and stochastic fluctuations, because that will bring too many subjects in a single paper.
Models that can reproduce the patterns of such stochastic fluctuations will be explored in the future.

\section{Behavioral Analysis of Experiment Results}\label{sec:ExperimentResults}

\subsection{Switching Rate}

To measure the switching behavior over rounds, we first define an index as follows.

{\bf Definition: Switching Rate}. {\it Suppose that there are N feasible routes connecting an OD pair. Given any two routes $i$ and $j$ in the feasible route set, the switching rate $p_{ij}^t$ is defined as the proportion of travelers switching from route $i$ to route $j$ during time $t$ and the next time $t+1$ ($t \geq 1$).} Obviously, $p_{ii}^t$ is the proportion of travelers remaining on their last-chosen route $i$ and $p_{ii}^t=1-p_{ij}^t$ if there are only two routes.

\subsection{Benchmark Models}

Three classic models are selected as the benchmarks of describing the experiment results, 
i.e., the Perfectly Rational Choice (PRC) model, the Multinomial Logit (MNL) model, and the Rational Behavior Adjustment Process (RBAP) model.

The PRC model is an intuitive, rational and individual level route choice rule.
It requires that ``{\it a traveler only changes his route to a less costly one, and, at least some travelers, if not all, will do so unless all the travelers were all on the least cost routes the previous day}" \citep{Zhang2001}.

\begin{figure}[!h]
    \centering
    \includegraphics[width=3in]{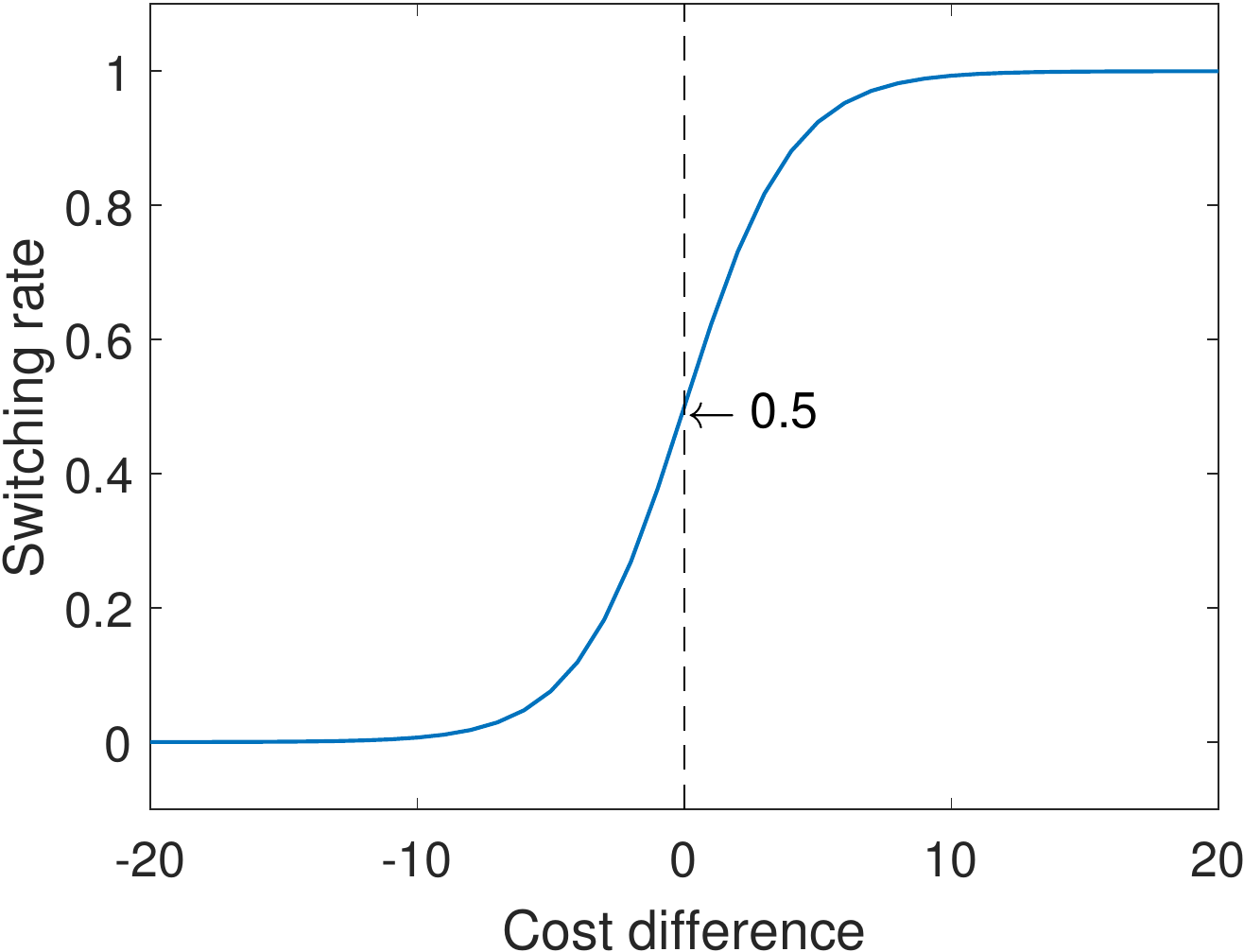}
    \caption{
    The switching rates predicted by the classic MNL model.} 
    \label{fig:MNL}
\end{figure}

The MNL model based on the RUM theory captures travelers' perception errors in making decision of route choices \citep{sheffi1985urban}.
Travelers are assumed to choose routes in a perceived (rather than actual) utility-maximum manner \citep{Daganzo1977}.
The MNL model can be simplified to a binary form for a simple network with two-parallel routes. 
Figure \ref{fig:MNL} presents the switching rates estimated by using the MNL model.
It can be seen that the switching rate is 0.5 at the DUE point where the cost of route $i$ is equal to that of route $j$. 
When the MNL model is applied as a route choice criterion, it leads to an SUE of network traffic at the aggregate level \citep{sheffi1985urban, cantarella2019dynamics}.

Different from the above two models, the RBAP model defines the changing patterns of route flow rather than individual route choice, and thus it is an aggregate model.
The RBAP model assumes that ``{\it the aggregated travel cost of the network decreases based on the previous day’s path travel costs when path flows change from day to day, and path flows become stationary over days if user equilibrium is reached}" \citep{Yang2009a}.
It is a relaxed assumption, compared with the strong assumption of the PRC model that requires all travelers to be perfectly rational \citep{Zhang2001}.
\cite{Yang2009a} showed that five popular DTD adjustment mechanisms are RBAP, making the RBAP model to be a representative model for the DTD dynamics.

Apparently, the PRC and MNL models were proposed based on different individual behavioral assumptions. 
Thus, it is interesting to see if their behavioral assumptions can satisfactorily explain the experimental observations; 
we will answer the question in Section \ref{Sec:observed_SR}.
In contrast, the RBAP model describes the aggregated-level flow adjustment process instead of travelers' choice.
Although it is interesting to test if the experimental data is consistent with the prediction of the RBAP model, we leave it to the future, since the focus of the paper is route choice behavior.

\subsection{Observed Switching Rate} \label{Sec:observed_SR}


This subsection provides an illustrative analysis on the observed switching rates, aiming to find gaps between the existing theories and our experimental observations.
The following two methods are employed to investigate the relationship between average switching rates and cost combinations.
Note that we only analyze the two-route scenarios here and leave the analysis of the three-route scenarios in Section \ref{sec:three-route}, since it is difficult to visually present the result of the more complicated three-route scenarios.

\begin{figure}[!htbp]
    \centering
    \subfigure[Scenario 1]{
    \includegraphics[width=3.3in]{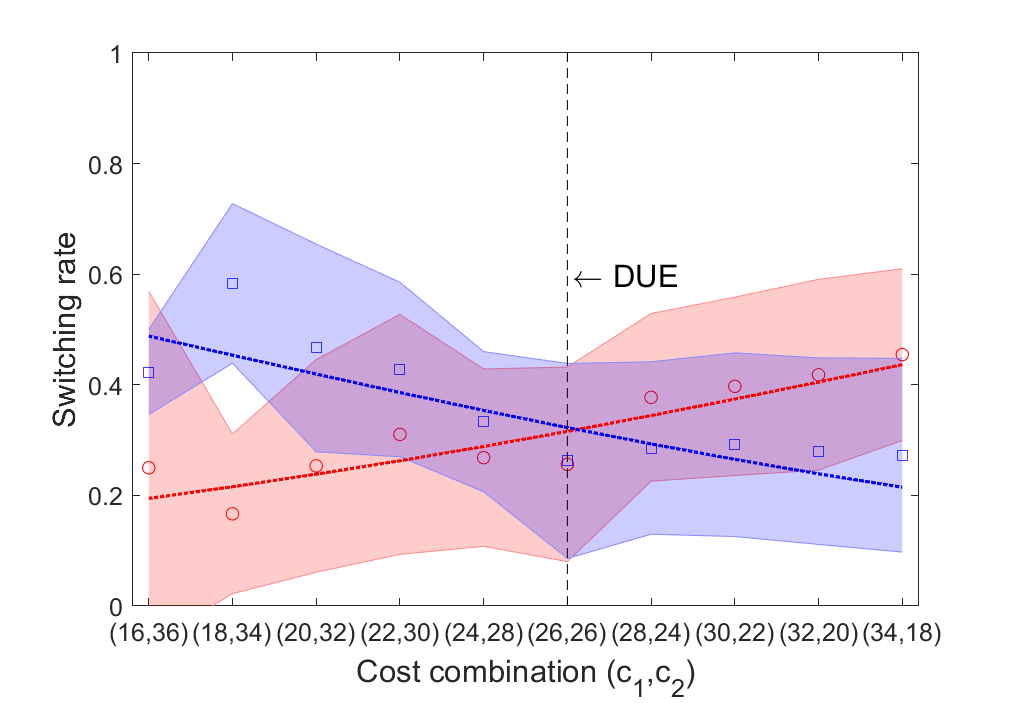}}
    \hspace{-8mm}
    \subfigure[Scenario 2]{
    \includegraphics[width=3.3in]{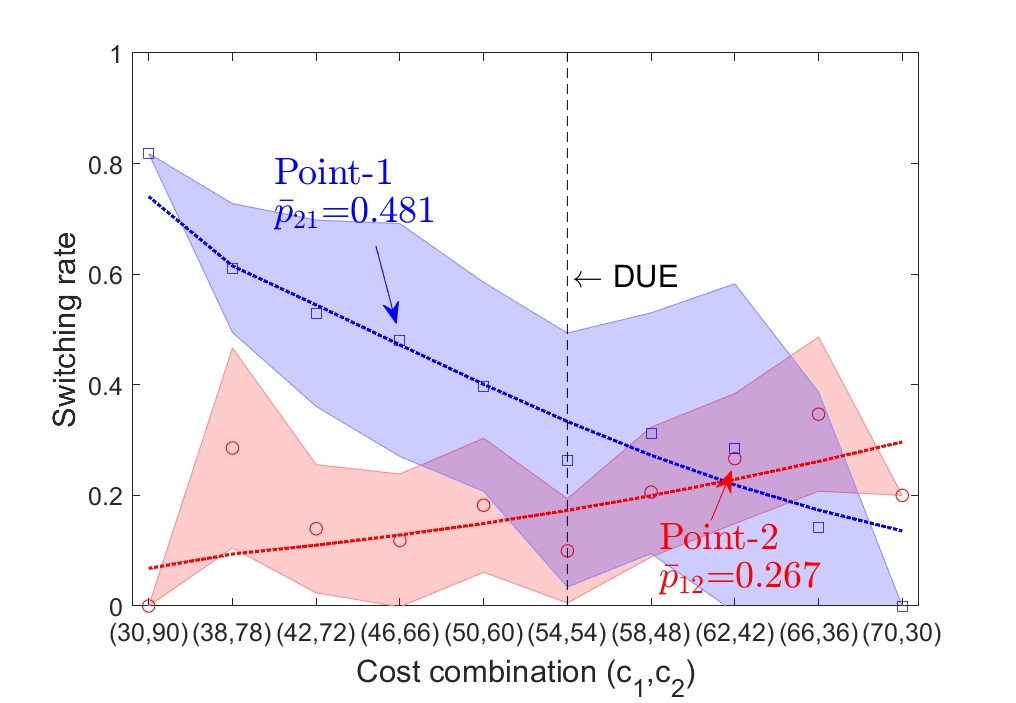}}
    \subfigure[Scenario 3]{
    \includegraphics[width=3.3in]{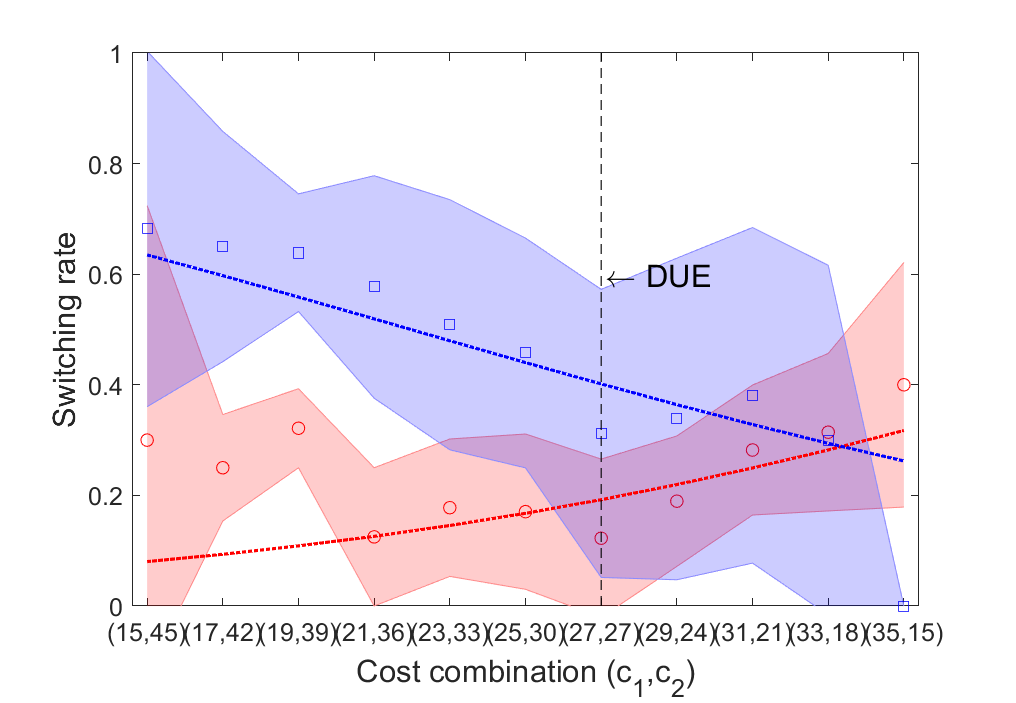}}
    \hspace{-8mm}
    \subfigure[Scenario 4]{
    \includegraphics[width=3.3in]{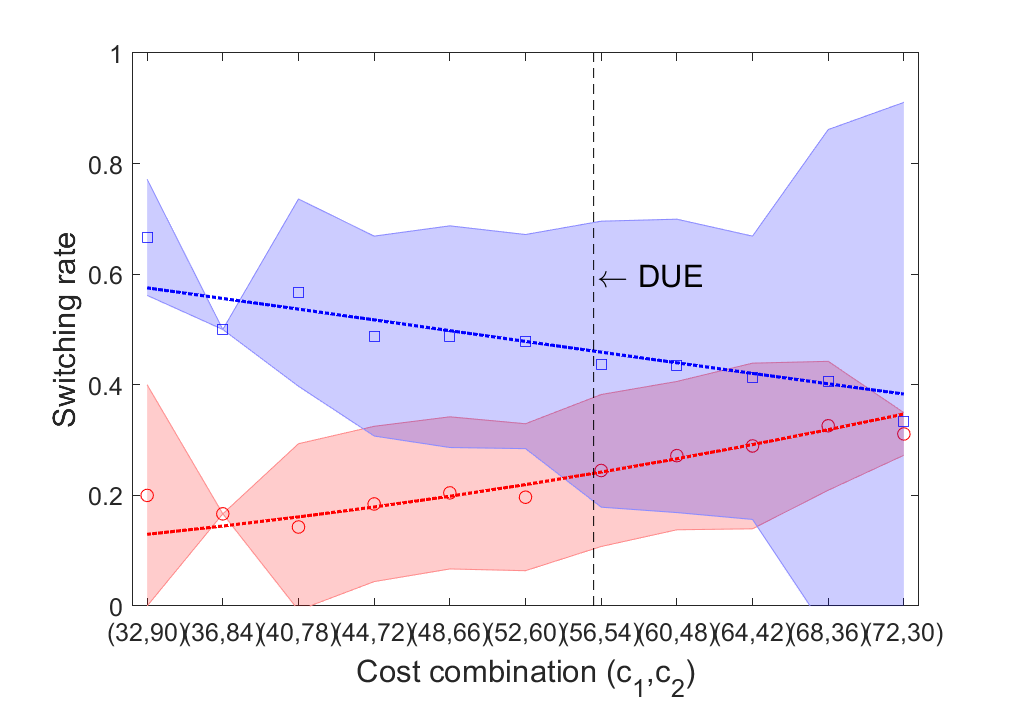}}
    \subfigure[Scenario 5]{
    \includegraphics[width=3.3in]{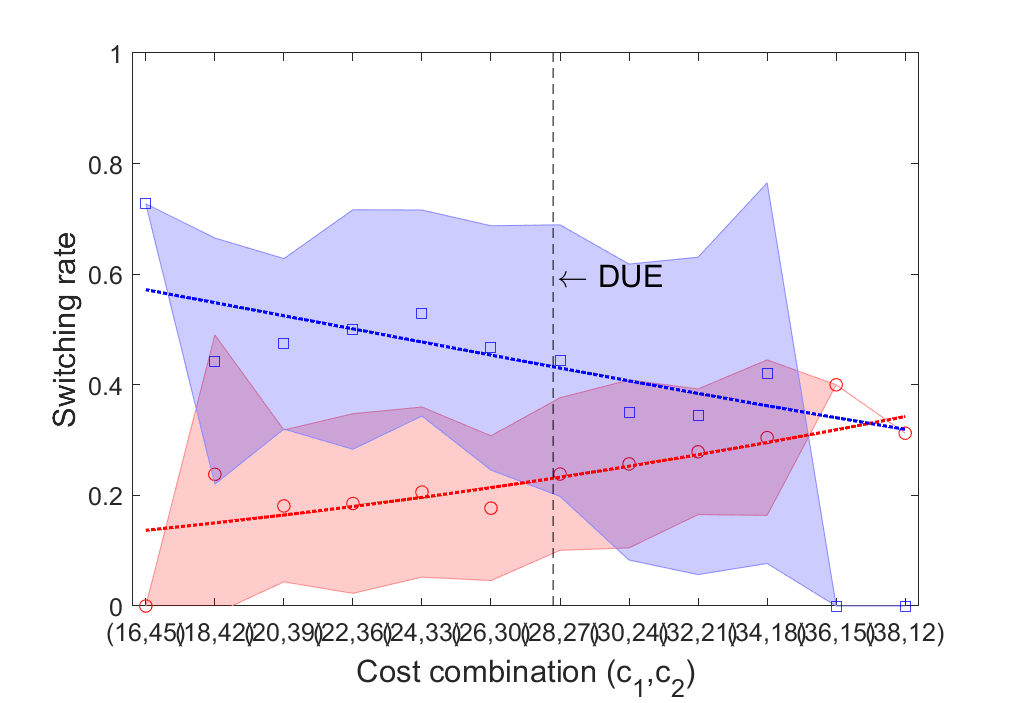}}    
    \hspace{-8mm}
    \subfigure{
    \includegraphics[width=3.3in]{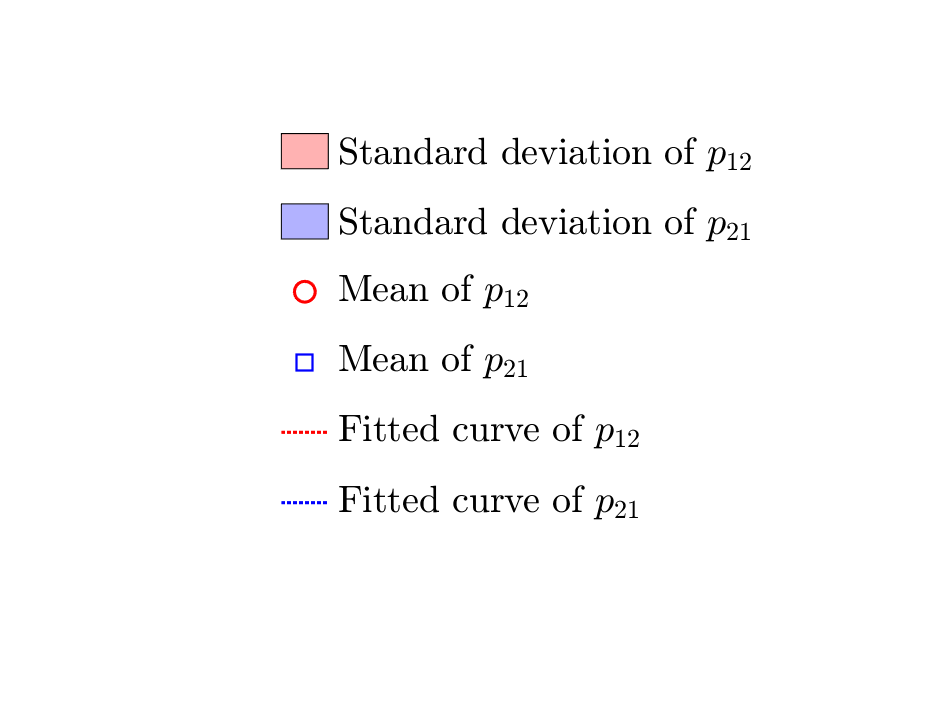}}    
    \caption{Observed switching rates for all the two-route scenarios (Scenarios 1-5).}
    \label{fig:Observed}
\end{figure}

\begin{figure}[!htbp]
    \centering
    \subfigure[Scenario 1]{
    \includegraphics[width=3.3in]{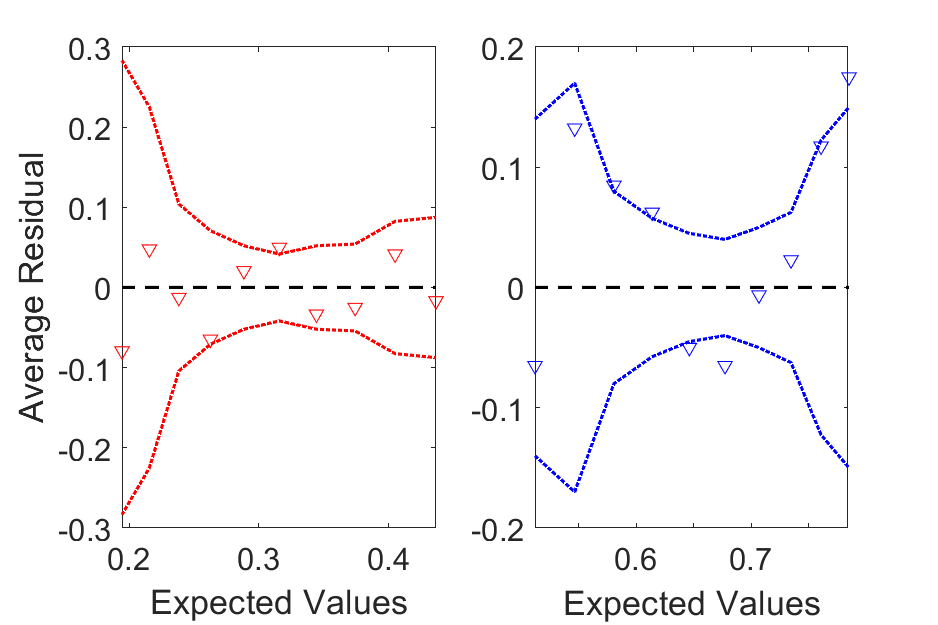}}
    \hspace{-8mm}
    \subfigure[Scenario 2]{
    \includegraphics[width=3.3in]{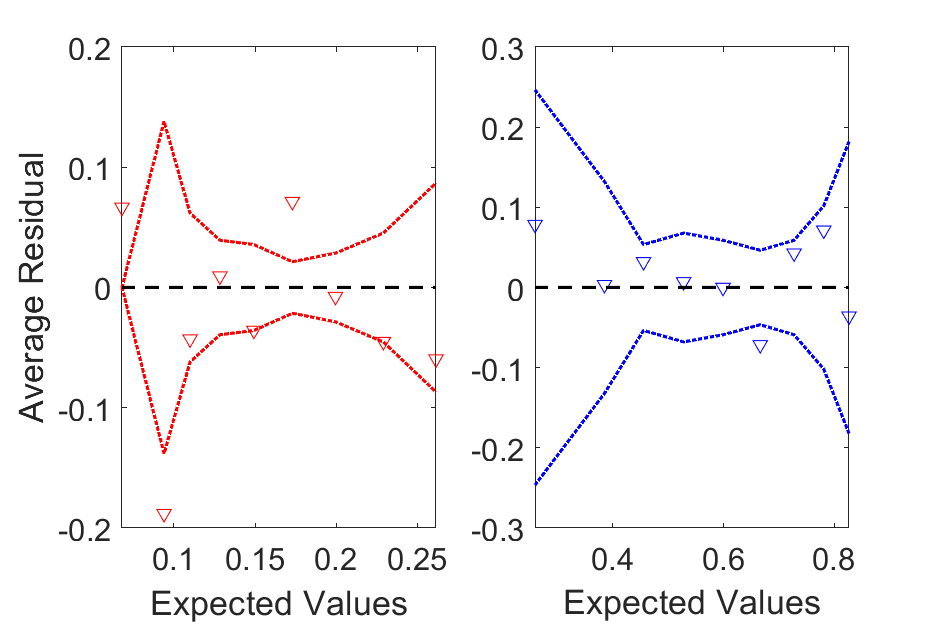}}
    \subfigure[Scenario 3]{
    \includegraphics[width=3.3in]{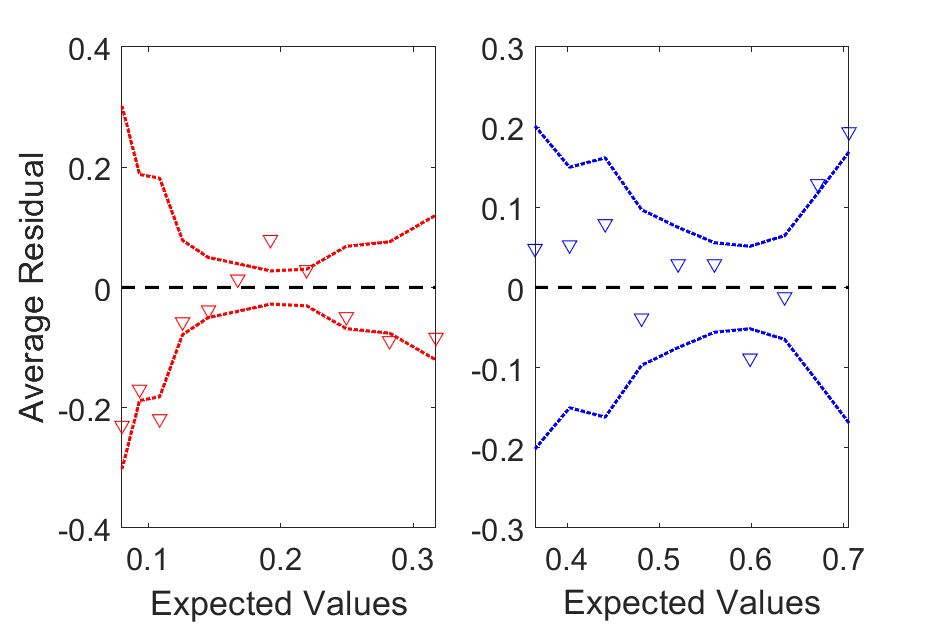}}
    \hspace{-8mm}
    \subfigure[Scenario 4]{
    \includegraphics[width=3.3in]{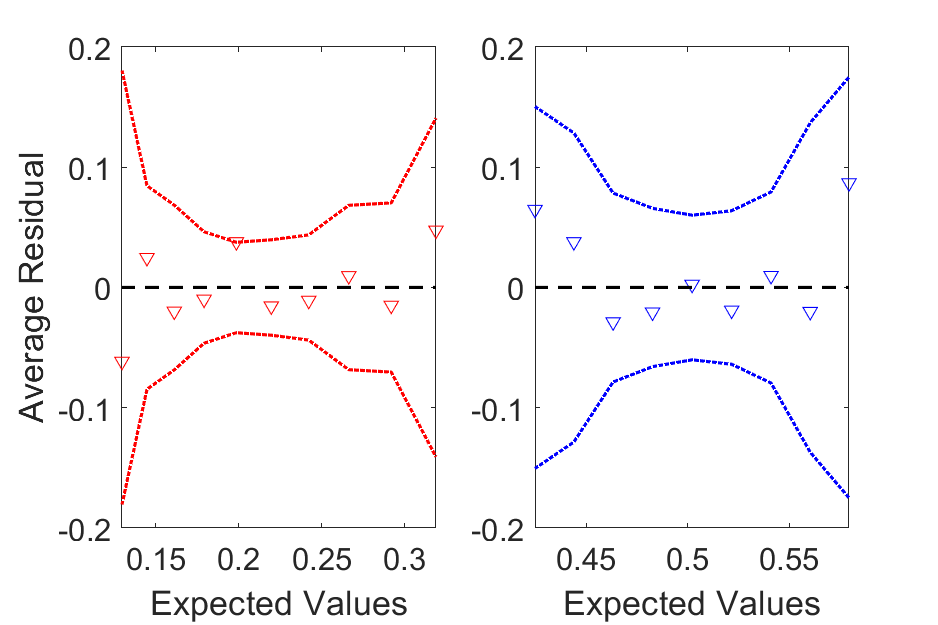}}
    \subfigure[Scenario 5]{
    \includegraphics[width=3.3in]{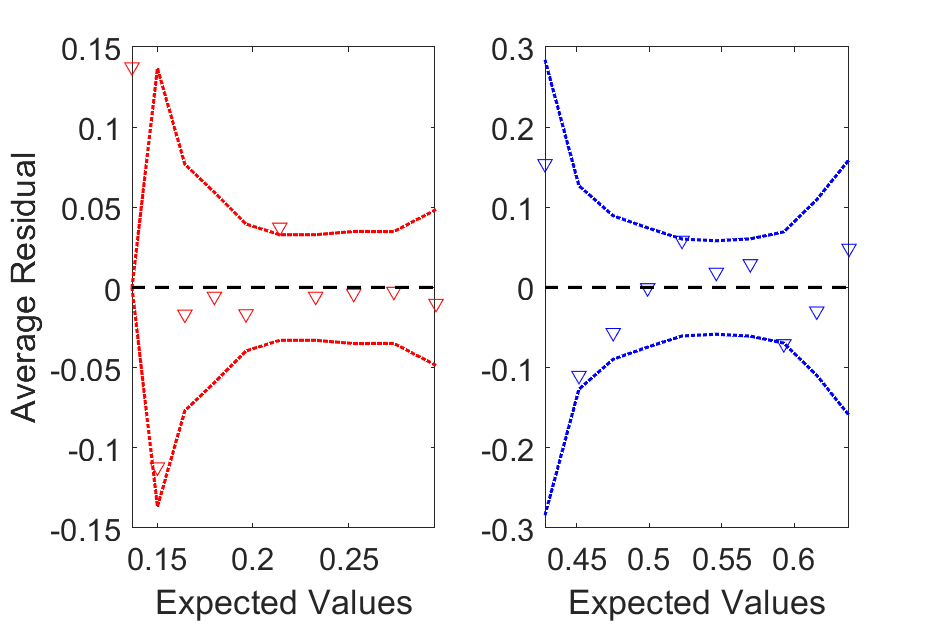}}    
    \hspace{-8mm}
    \subfigure{
    \includegraphics[width=3.3in]{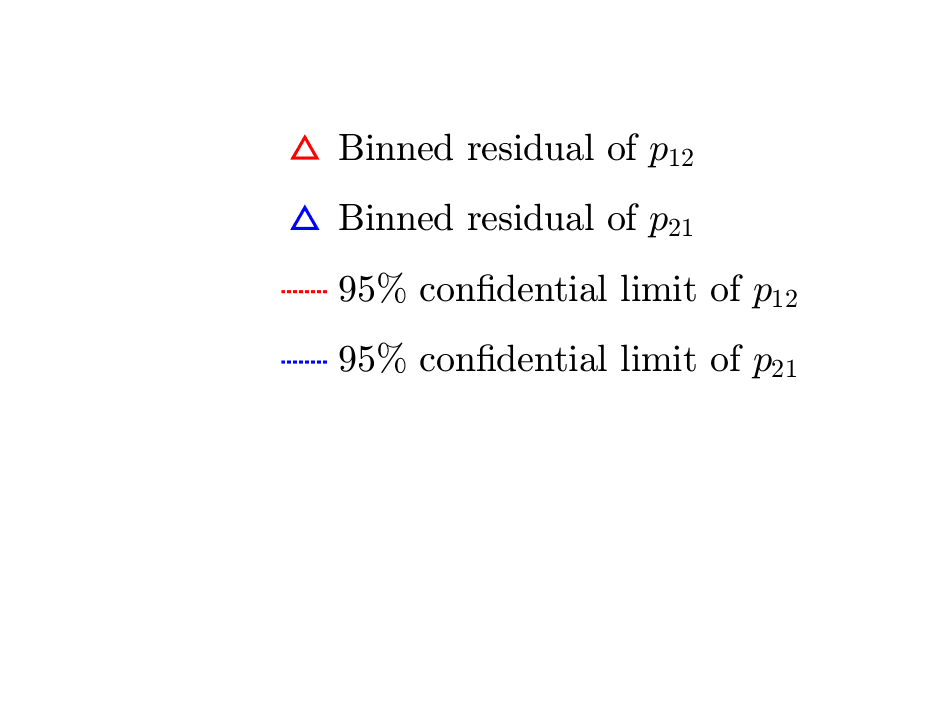}}    
    \caption{Binned residual plots for all the two-route scenarios (Scenarios 1-5).}
    \label{fig:Residual}
\end{figure}

\begin{itemize}
\vspace{-3 mm}
    \setlength{\itemsep}{0pt}
    \setlength{\parsep}{0pt}
    \setlength{\parskip}{0pt}

\item {\bf Method 1}. We employ average switching rate $\bar{p}_{ij}$ as an intuitive indicator, which is calculated as follows. 
\begin{equation}
    \bar{p}_{ij}(\vec{c})= \frac{\sum_{t\in T(\vec{c})}
    p_{ij}^t}{|T(\vec{c})|}
\end{equation}
where $\vec{c}$ is a specific cost combination and $T(\vec{c})$ is a set of the rounds when the costs of all routes are equal to $\vec{c}$. 
The average switching rate can be treated as a reflection of the average behavior of the population with the elimination of individual heterogeneity and stochastic factors. The results are presented in Figure \ref{fig:Observed}.

\item 
{\bf Method 2}. 
We apply the Logistic regression to describe the binary-choice process in the two-route scenarios. 
The model of the switching rate, which is written as follows, is fitted by using the maximum likelihood estimation and all experimental data.
\begin{equation}
    p_{ij} = \frac{1}{1+e^{-(\theta_0+\theta_1(t_i-t_j))}}
\end{equation}
where the standard form of the Logistic regression is directly employed, since we intend to explain the data by fitting a statistical model before formally proposing a model from the choice behavioral perspective.
In comparison to the simple averaging strategy, the contribution of data corresponding to different cost combination could be appropriately weighted through maximum likelihood estimation, regardless of the frequency.
The fitted curve is presented in Figure \ref{fig:Observed} with the parameters shown in Table \ref{tab:para} where the results indicate that all variables are significant. 
The fit of the model is further assessed by using binned residual plots \citep{gelman2006data} as shown in Figure \ref{fig:Residual}. 
It can be seen that there is no systematic patterns and most data points fall between the 95\% confidential limits, indicating that the Logistic regression model can effectively describe the trend of the experimental data.
\end{itemize}

\begin{table}[!htbp] 

\footnotesize\center  
\setlength{\tabcolsep}{7mm}{
\caption{Regression results of switching rates for all the two-route scenarios (Scenarios 1-5).} \label{tab:para}}

\begin{tabular}{ccccccc}
\toprule
Scenario & & \multicolumn{2}{c}{$p_{12}$} & & \multicolumn{2}{c}{$p_{21}$}  \\
         \cline{3-4} \cline{6-7}  
         & & Coef. & $p$-value & & Coef. & $p$-value\\
\midrule
 1 & $\theta_1$ & 0.0324 &0.000***& & 0.0347 &0.000*** \\
   & $\theta_0$ & -0.773 &0.000*** & & -0.741 &0.000*** \vspace{2mm}\\
 2 & $\theta_1$ & 0.0175 &0.000*** & & 0.0290 &0.000*** \\
   & $\theta_0$ & -1.564 &0.000*** & & -0.692 &0.000*** \vspace{2mm}\\
 3 & $\theta_1$ & 0.0334 &0.000*** & & 0.0318 &0.000*** \\
   & $\theta_0$ & -1.435 &0.000*** & & -0.399 &0.000*** \vspace{2mm}\\
 4 & $\theta_1$ & 0.0127 &0.000*** & & 0.00779 &0.005** \\
   & $\theta_0$ & -1.166 &0.000*** & & -0.149 &0.008** \vspace{2mm}\\
 5 & $\theta_1$ & 0.0216 &0.000*** & & 0.0190 &0.000*** \\
   & $\theta_0$ & -1.214 &0.000*** & & -0.262 &0.000*** \vspace{2mm}\\
\bottomrule
\end{tabular}
\begin{tablenotes}
 \footnotesize
\item[] \ \  *$p<$0.05, \ \ **$p<$0.01, \ \ ***$p<$0.001
 \end{tablenotes} 
\end{table}

We have the following observations and remarks regarding the experimental switching rate from the results presented in Figure \ref{fig:Observed}.

\begin{itemize}
\vspace{-3 mm}
    \setlength{\itemsep}{0pt}
    \setlength{\parsep}{0pt}
    \setlength{\parskip}{0pt}

\item
{\bf Remark 1}: {\it The switching rate has an increasing trend with the cost difference between the last-chosen and its alternative routes; The switching rate is still positive even if the last-chosen route is better than the alternative route.}

{\bf Corresponding observations}.
It can be seen from Figure \ref{fig:Observed} that the average switching rates $\bar p_{21}$ (blue squares) decrease with the increase of the value of $c_2-c_1$ (i.e., the left-to-right direction of the axis of cost combination), indicating that {\it the switching rate increases with the cost difference between the last-chosen and its alternative routes}.
The average switching rate is still larger than zero when $c_1<c_2$ (i.e., the left part of the DUE line), indicating that {\it the switching rate is still positive even if the last-chosen route is better than the alternative route.}

\ \ \ \ \ The PRC model cannot describe this phenomenon, since it assumes that travelers will never switch to any routes with higher cost \citep{Zhang2001}. 
In contrast, the MNL model allows travelers to switch routes even if their last choices are actually better, and thus it is more consistent with the observation. 

\item {\bf Remark 2}: {\it When the network is asymmetric, the average switching rates of both routes at the DUE point are between 0 and 0.5, and they are significantly different from each other.}

{\bf Corresponding observations}. In Figure \ref{fig:Observed}(b)-(e), i.e., the asymmetric scenarios, the average switching rates of route 2 (blue squares) are all above those of route 1 (red circles) at the DUE points. 
In contrast, the blue square coincides with the red circle at the DUE point in Figure \ref{fig:Observed}(a), i.e., the symmetric scenario.
In Table \ref{tab:para}, it can be also found that all $\theta_0$ of $p_{21}$ are larger than those of $p_{12}$ in the asymmetric scenarios, meaning that the switching probabilities from route 2 to route 1 at DUE ($t_1-t_2=0$) are larger than those from route 1 to route 2.



\ \ \ \ \ The \textit{t}-test is employed to evaluate if the above observations are statistically significant. 
The results in Table \ref{tab:Remark2} show that the observations in Remark 2 are supported at the significance level of 0.05. 
In contrast, the theoretical switching rates predicted by the PRC and MNL models are 0 and 0.5, respectively, turning out that neither the MNL nor PRC models could explain those observations.

\begin{table}[!htbp]\footnotesize\center
\setlength{\tabcolsep}{5mm}{
\caption{Hypothesis tests for Remark 2 in all the two-route scenarios (Scenarios 1-5).}
\label{tab:Remark2}
\begin{tabular}{llllll}
\hline
Alternative &
\multicolumn{5}{c}{$p$-value}\\
\cline{2-6} 
Hypothesis & Scenario 1         & Scenario 2         & Scenario 3         & Scenario 4         & Scenario 5         \\ 
\hline
$\bar{p}_{12}<0.5$  \ at DUE        & 0.000***     & 0.000***      & 0.000***     & 0.000***     & 0.000***            \\
$\bar{p}_{21}<0.5$ \ at DUE        & 0.000***     & 0.000***      & 0.000***     & 0.034*     & 0.036*            \\
$\bar{p}_{12} \neq \bar{p}_{21}$ \ at DUE    & 0.832   & 0.000***     & 0.000***     & 0.000***     & 0.000***            \\
\hline
\end{tabular}
\begin{tablenotes}
 \footnotesize
\item[]  *$p<$0.05, \ \ **$p<$0.01, \ \ ***$p<$0.001
 \end{tablenotes} }
\end{table}

\item 
{\bf Remark 3}: {\it The average switching rate depends not only on the cost difference of routes but also on the characteristics of the last-chosen route.}

\noindent {\bf Corresponding observations}. Take two cost combinations in Scenario 2 as examples and see the labeled Point-1 and Point-2 in Figure \ref{fig:Observed}(b).
It can be found that
\begin{itemize}
\vspace{-3 mm}
    \setlength{\itemsep}{0pt}
    \setlength{\parsep}{0pt}
    \setlength{\parskip}{0pt}

\item {\it Point-1}: The average switching rate $\bar p_{21}=0.481$ when the cost of the last-chosen route (i.e., route 2) is 20 higher than that of its alternative route (i.e., route 1), i.e., $(c_1, c_2)=(46,66)$ and $c_2-c_1=20$.

\item {\it Point-2}: The average switching rate $\bar p_{12}=0.267$ when the cost of the last-chosen route (i.e., route 1) is 20 higher than that of its alternative route (i.e., route 2), i.e., $(c_1, c_2)=(62,42)$ and $c_1-c_2=20$.
\end{itemize}

\ \ \ \ \ Table \ref{tab:Remark3} presents all switching rates and the statistical test results when the cost difference of routes are the same in Scenarios 1-3.
The two-proportion \textit{z}-test is carried out to test if the observations are statistically significant.
For Scenario 1, the test results indicate that no significant difference exists between $\bar p_{12}$ and $\bar p_{21}$ for all cases under which the cost differences of routes are the same at the significant level of 0.05. 
In contrast, the alternative hypothesis ``$p_{21} \neq p_{12} $" is accepted in the most cases in Scenarios 2 and 3 with asymmetric networks, meaning that the switching rates are different even when the cost differences of routes are the same in these two scenarios.
Two exceptions are found, in both of which there exists one route with a small number of travelers (e.g. $c_2=36$ in Scenario 2 turns out that only two travelers are on route 2 derived from the cost function $c_2=6f_2+24$; see Table \ref{tab:Parameters}).
This leads to a large switching rate deviation and makes the results not predictable. 

\end{itemize}

\begin{table}[!htbp]\footnotesize\center
\setlength{\tabcolsep}{3.7mm}{
\caption{The statistical tests of the different switching rates when the cost difference of routes are the same.}\label{tab:Remark3}
\begin{tabular}{ccccccccccll}
\toprule
Scenario & \multicolumn{4}{c}{route 1 $\rightarrow$ route 2} & & \multicolumn{4}{c}{route 2 $\rightarrow$ route 1} & \multicolumn{2}{c}{$p$-value}\\
\cline{2-5} \cline{7-10}  
  & $c_1$ & $c_2$ & $c_1-c_2$ & $\bar{p}_{12}$ & & $c_2$ & $c_1$ & $c_2-c_1$ & $\bar{p}_{21}$ &  (H1: $\bar{p}_{21}\neq \bar{p}_{12}$) \\
\midrule
1 & 20 & 32 & -12 & 0.253 & & 20 & 32 & -12 & 0.353 & \ \ \ \ \ \ 0.901\\
  & 22 & 30 & -8 & 0.310 & & 22 & 30 & -8 & 0.292 & \ \ \ \ \ \ 0.773\\
  & 24 & 28 & -4 & 0.267 & & 24 & 28 & -4 & 0.286 & \ \ \ \ \ \ 0.693\\
  & 28 & 24	& 4 & 0.377 & & 28 & 24 & 4 & 0.305 & \ \ \ \ \ \ 0.222\\
  & 30 & 22 & 8 & 0.397 & & 30 & 22 & 8 & 0.428 & \ \ \ \ \ \ 0.483  \\
  & 32 & 20 & 12 & 0.363 & & 32 & 20 & 12 & 0.503 & \ \ \ \ \ \ 0.504  \vspace{2mm}\\
\midrule
2 & 42 & 72 & -30 & 0.154 & & 36 & 66 & -30 & 0.142  & \ \ \ \ \ \ 0.709\\
  & 46 & 66 & -20 & 0.119 & & 42 & 62 & -20 & 0.286  & \ \ \ \ \ \ 0.000***\\
  & 50 & 60 & -10 & 0.182 & & 48 & 58 & -10 & 0.313  & \ \ \ \ \ \ 0.000***\\
  & 58 & 48 & 10 & 0.206 & & 60 & 50 & 10 & 0.396 &  \ \ \ \ \ \ 0.000***\\
  & 62 & 42 & 20 & 0.266 & & 66 & 46 & 20 & 0.481 &  \ \ \ \ \ \ 0.000*** \\ 
  & 66 & 36 & 30 & 0.346 & & 72 & 42 & 30 & 0.574 &  \ \ \ \ \ \ 0.008** \vspace{2mm}\\
\midrule
3 & 21 & 36 & -15 & 0.125 & & 18 & 33 & -15 & 0.333  &  \ \ \ \ \ \ 0.045*\\
  & 23 & 33 & -10 & 0.178 & & 21 & 31 & -10 & 0.458  &  \ \ \ \ \ \ 0.001**\\
  & 25 & 30 & -5 & 0.171 & & 24 & 29 & -5 & 0.338 &  \ \ \ \ \ \ 0.000***\\
  & 29 & 24 & 5 & 0.190 & & 30 & 25 & 5 & 0.457 &  \ \ \ \ \ \ 0.000***\\
  & 31 & 21 & 10 & 0.298 & & 33 & 23 & 10 & 0.508  & \ \ \ \ \ \ 0.000*** \\
  & 33 & 18 & 15 & 0.367 & & 36 & 21 & 15 & 0.576  & \ \ \ \ \ \ 0.000***\\
\bottomrule
\end{tabular}}
\begin{tablenotes}
 \footnotesize
\item[] \ \  *$p<$0.05, \ \ **$p<$0.01, \ \ ***$p<$0.001
 \end{tablenotes} 
\end{table}

In summary, the above analyses suggest that both the RBAP and MNL models, which only take the cost difference into account, cannot completely describe the observed experiment results. 
In particular, Remarks 1 and 2 imply that travelers make route choices in a Logit-like way with ``inertia", i.e., travelers prefer to stick to the last-chosen route. 
Furthermore, Remark 3 indicates that travelers' choice behavior is influenced not only by the previous route costs but also by other characteristics of their last-chosen routes, implying that travelers treat routes with different inherent characteristics differently. 
To fully reflect the experimental observations, we will propose an analytical DTD model in Section \ref{sec:Model}.

\section{Analytical DTD Dynamic Model}\label{sec:Model}

\subsection{Behavior Assumptions of Route-dependent Attractions}

As shown in Remark 1, the MNL model performs better than the PRC model.
Therefore, we choose the RUM-based model (the Logit choice model, specifically) as the foundation of proposing new models.

Then, we define {\it route-dependent attractions} to explain the observations shown in Remarks 2 and 3, i.e., subjects treat routes differently. 
To that end, we separate the route-choice decision process into two steps, which are dominated by the following two route-dependent attractions, respectively.

\begin{itemize}
\vspace{-3 mm}
    \setlength{\itemsep}{0pt}
    \setlength{\parsep}{0pt}
    \setlength{\parskip}{0pt}

\item
{\it Route-dependent inertia}. 
Inertia means that travelers have a constant tendency to stay on their last-chosen route, regardless of the cost of the route in the last round. 
More importantly, the inertia here is route-dependent, indicating that the tendency for travelers diverse across routes. 

\item
{\it Route-dependent preference}. 
The travelers who intend to break the inertia will re-consider their route choices.
The switching decision is affected not only by the cost difference between their last-chosen and alternative routes, but also by traveler's route attribute-related preference. 

\end{itemize}

There would be various psychological mechanisms or theoretical basis underlying the above assumptions. 

\begin{itemize}
\vspace{-3 mm}
    \setlength{\itemsep}{0pt}
    \setlength{\parsep}{0pt}
    \setlength{\parskip}{0pt}
    
    \item 
    We use the term ‘inertia’ mostly to describe the tendency of status quo maintenance (i.e., sticking to the previous choice). Multiple possible psychological mechanisms may explain such status-quo-maintenance behavior tendency \citep{Qi2019}, including ‘indifference band’ \citep{Hu1997}, ‘effort-accuracy trade-offs’ \citep{Chorus2012},  ‘regret-aversion’ \citep{Chorus2008}, or simply  human laziness, etc. It may also result from 'strategic thinking', that is, people decide to stay because of their belief that other people on their choice will switch and lead their choice to be less crowded.
    
    \item 
    A status-quo-maintenance category (equal to players with route-dependent ‘inertia’ in this paper), and a highly-risk-averse category (roughly equal to players with route-dependent ‘preference’) have been observed by \citet{Qi2019}. Based on the previous findings, we believe that one of the explanations for route-dependent preference lies in risk aversion. According to the well-known mean-variable decision rule \citep{Meyer1989,Noland2002,Avineri2005a}, one makes a decision not only based on the mean expected utility of a candidate option but also by taking the variance of the option into account, as most people are risk-averse to large variance.

\end{itemize}

Similar terminologies were previously discussed in several studies, while the definitions here are different from the existing ones in the following aspects.

\begin{itemize}
\vspace{-3 mm}
    \setlength{\itemsep}{0pt}
    \setlength{\parsep}{0pt}
    \setlength{\parskip}{0pt}
    
    \item 
    Choice inertia was used in route choice modeling \citep{Zhang2015a, Cantarella2016}. In \cite{Zhang2015a}, inertial is defined as travelers choosing routes from a subset of all alternative routes, which is different from the definition in this paper. 
    The definition here is similar to that in \cite{Cantarella2016}, while we find that inertia is route-dependent and we provide experimental evidence.
    
    \item     
    A hypothesis of route-dependent preference was proposed and tested by using virtual experimental data by \cite{Ye2018}. 
    It assumed that the flow swapping among routes depends on the origin and target routes as well as the cost difference, and no empirical data was provided to support the hypothesis.
    Differently, the route-dependent preference in this paper is assumed to influence traveler's route switching probabilities, which provides a better behavioral explanation of route-dependent preference.
    Moreover, it is supported by the experimental observation.
    
    \item     
    More importantly, we simultaneously integrate the inertia and route-dependent preference into a discrete choice-based analytical model, which can satisfactorily reproduce the experimental observations. It is one of the main contributions of the paper.
    
\end{itemize}




\subsection{Route Switching Model}

As some existing studies did, we consider only a network with one OD pair connected by multiple routes. 
The simple network could help focus on the choice procedure instead of other factors that are not concerned in the paper.
The expansion of multiple OD pairs will be left for the future.

According to the assumption of the route-dependent attractions, we propose that the route $i$ has a route-specific and constant {\it attraction coefficient} denoted by $\eta_i$. 
Assume that travelers in the same scenario are homogeneous, and thus route $i$ has the same attraction for all travelers. 
The switching rate $p^t_{ij}$ from the currently used route $i$ to any other route $j$ at time $t$ could be calculated as follows.
\begin{equation}\label{equ:pji}
    p^t_{ij}=
    \begin{cases}
    \begin{split}
    &P_i\hat p^t_{ij}\ , \ & \ \text{if}\ \ i\neq j\\
    &(1-P_i) + P_i\hat p^t_{ij}\ , \ & \ \text{if}\ \ i=j
    \end{split}
    \end{cases}
\end{equation}
where 
$P_i$ ($P_i\in[0,1]$) is the proportion of the travelers who re-consider their route choices on route $i$. $P_i$ might be related to many factors such as the characteristics of other routes. However, to make the model analyzable we simply assume that $P_i$ is a constant that only relates to the property of route $i$, such as route length and speed limit, and we will show that the model has satisfactory performance over the experimental data.
$\hat p^t_{ij}$ ($\hat p^t_{ij}\in[0,1]$) is the final chosen rate, i.e., the proportion of the travelers who switch to route $j$ in the proportion of $P_i$. 
More specifically, we explain Equation \ref{equ:pji} as follows.

The term $(1-P_i)$ indicates the route-dependent inertia behavior, i.e., only a proportion of travelers ($P_i$) will re-consider their choices and the remaining proportion ($1-P_i$) simply sticks to the last-chosen route. 
Let $P_i$ be a decreasing function of $\eta_i$ (written as $P_i=P(\eta_i)$), considering the fact that less travelers will re-consider to leave higher-attraction routes. 

The utility $u_i^t$ of choosing route $i$ at time $t$ is defined to be the sum of an additive function of a deterministic component and a random error term $\zeta$.
\begin{equation}
    u_i^t = -C_i^t + \zeta 
\end{equation}
where $C_i^t=C(c_i^t,\eta_i)$ and $c_i^t$ is the cost of route $i$ at time $t$; 
$C(c_i^t,\eta_i)$ is the generic cost function with the following considerations. 
Choosing the lower-cost route leads to higher utility.
For the same cost routes, choosing the route with a higher attraction coefficient will get higher utility. 
It turns out that
${\partial C}/{\partial c_i^t}>0$
and ${\partial C}/{\partial \eta_i}<0$.
Therefore, the route-dependent preference is reflected by $C_i^t$. 

In the existing studies, the cost updating process is suggested as travelers updating their perception or routes according to the history travel costs \citep{cantarella1995dynamic, Cantarella2016, cantarella2019dynamics}. 
Differently, this study assumes that travelers make decisions based on only the costs of the last round. This setting helps focus on the main assumptions (route-dependent attractions), and it is consistent with many existing models.
The role of cost updating is an important problem that is worth to be explored in the future.

Moreover, assume that the random error term $\zeta$ follows the independently and identically distributed (IID) Gumbel distribution, and then  
\begin{equation}\label{equ:bar_pji}
    \hat p^t_{ij} = \frac{e^{-\theta C^t_j}}{\sum_{k=1}^N e^{-\theta C^t_k}}
\end{equation}
where $\theta$ is a dispersion parameter.

Finally, the switching rate $p^t_{ij}$ from route $i$ to route $j$ at time $t$ is proposed as follows by substituting Equation \ref{equ:bar_pji} into Equation \ref{equ:pji}.
\begin{equation}\label{equ:pji_final}
    \text{Switching Rate: ~~~} p^t_{ij}=
    \begin{cases}
    \begin{split}
    &P_i \frac{e^{-\theta C_j^t}}{\sum_{k=1}^N e^{-\theta C_k^t}}\ , \ & \ \text{if}\ \ i\neq j\\
    &(1-P_i) + P_i\frac{e^{-\theta C_j^t}}{\sum_{k=1}^N e^{-\theta C_k^t}}\ , \ & \ \text{if}\ \ i=j
    \end{split}
    \end{cases}
\end{equation}
with the route-dependent inertia $P_i=P(\eta_i)$ and the route-dependent preference $C_i^t=C(c_i^t,\eta_i)$.

The above choice procedure is suitable to deal with multiple routes between an OD pair, since travelers on route $i$ would generate a switching rate to every other route $j$ between the same OD pair in each round.
Essentially, $p^t_{ij}$ is a conditional route choice rate depending on both referenced and target routes \citep{Ben-Akiva1990,cantarella2019dynamics}.

\subsection{Network Flow Evolution Model}

Suppose that there are $f_i^t$ travelers on route $i$ at time $t$ (i.e., route flow).
The  expected value of flow switching from route $i$ to route $j$, which is denoted by $\Delta f_{ij}^t$, is written as follows. 
\begin{equation}\label{equ:Delta_h}
    \Delta f_{ij}^t= p_{ij}^t f_i^t
\end{equation}
Then, the flow on route $i$ at time $t+1$ is 
\begin{equation}\label{equ:hit+1}
    f_i^{t+1}=\sum_{k=1}^N \Delta f_{ki}^t
\end{equation}
where $N$ is the size of the feasible set of the routes connecting the same OD pair.

The flow on route $i$ at time $t$ can be obtained as follows by substituting Equation \ref{equ:pji_final} into Equation \ref{equ:hit+1}.
\begin{equation}\label{equ:hit+1-hit}
\begin{split}
    f_i^{t+1} = (1-P_i) f^t_i  +  e^{-\theta C_i^t}\frac{\sum_{k=1}^N P_k f^t_k}{\sum_{k=1}^N e^{-\theta C_k^t}} 
    \ \ \ \Rightarrow \ \ \  f_i^{t+1} - f_i^{t} = -P_i f^t_i  +  e^{-\theta C^t_i}\frac{\sum_{k=1}^N P_k f^t_k}{\sum_{k=1}^N e^{-\theta C^t_k}}.  
\end{split}
\end{equation}
A continuous-in-time form of Equation \ref{equ:hit+1-hit} can be written as follows.
\begin{equation}\label{equ:citform}
    \dot f_i = -P_i f_i  +   \sum_{k=1}^N P_k f_k\frac{e^{-\theta C_i}}{\sum_{k=1}^N e^{-\theta C_k}}.
\end{equation}
%
The model in Equation \ref{equ:citform} fits into the framework in \cite{cantarella1995dynamic} and it could also be seen as an extension of the exponential smoothing filter in \cite{Cantarella2016}. 
However, the differences between the proposed model and the existing ones are significant. 
First, the importation of the route-dependent attraction (in particular, the route-dependent inertia and preference) is enlightened by the observations from multi-scenario experimental data, endowing the model with the more explanatory power of human travelers’ route choice behaviors. 
Second, the route-dependent inertia makes the form of the proposed model different from that in \cite{Cantarella2016}, which requires different techniques for analyzing the characteristics of the model.

Let $\dot f_i=0$, and the fixed point of the flow evolution model, i.e., the equilibrium point of the network flow, is written as follows.
\begin{equation}\label{equ:hi_1}
    f_i = \frac{\sum_{k=1}^N P_k f_k}{P_i} 
    \frac{e^{-\theta C_i}}{\sum_{k=1}^N e^{-\theta C_k}}.
\end{equation}

It can be seen from Equation \ref{equ:hi_1} that the equilibrium of the model is a generalization of SUE, and it would be identical to SUE if $P_i=1$ and $C_i=c_i$. We will illustrate the equilibrium state of the model through numerical examples in Section 6.

\subsection{Existence and Uniqueness of the Equilibrium}

For an OD pair with $N$ routes, we formulate a minimization problem as follows.
\begin{equation}
    \begin{split}
    \min z(\vec f) \\
    \text{s.t.}  \ \ \ \  F=\sum_i^N f_i
    \end{split}
\end{equation}
where 
\begin{equation}
z(\vec f) = \sum_{i=1}^N  \theta\int_0^{f_i} C(c_i(x), \eta_i)\text{d}x
         + \sum_{i=1}^N  f_i  \text{ln}(P_i f_i)
\end{equation}
where $c_i(\cdot)$ is the flow-cost function of route $i$ and $F$ is the OD traffic demand. 
It is a convex programming problem and a unique minimum value can be obtained by using the Karush-Kuhn-Tucker conditions as follows.
Let 
\begin{equation}
    H= z(\vec f) - \lambda \left( \sum_{i=1}^N f_i - F \right)
\end{equation}
and then
\begin{equation}
    \frac{\partial H}{\partial f_i} 
    = \theta C_i + \left[ \text{ln} (P_i f_i) + 1 \right] - \lambda 
    =0 .
\end{equation}

With the assumption of the convex separable cost functions with respect to route flow,  for $\forall i,j\ (i\neq j)$, we have
\begin{equation}
 \theta C_i +  \left[ \text{ln} (P_i f_i) + 1 \right] 
= \theta C_j + \left[ \text{ln} (P_j f_j) + 1 \right] 
\end{equation}
and 
\begin{equation}\label{equ:hi_2}
    \begin{split}
    & \text{ln}(P_i f_i) - \text{ln}(P_j f_j) = \theta C_j - \theta C_i \\
    \Rightarrow \ \ \ \ \ \  & \frac{P_i f_i}{P_j f_j} = e^{\theta C_j - \theta C_i} = \frac{e^{-\theta C_i}}{e^{-\theta C_j}} \\
    \Rightarrow \ \ \ \ \ \  & \frac{\sum_{k=1}^N P_k f_k}{P_i f_i} = \frac{\sum_{k=1}^N e^{-\theta C_k}}{e^{-\theta C_i}} \\
    \Rightarrow \ \ \ \ \ \  & f_i = \frac{\sum_{k=1}^N P_k f_k}{P_i}  \frac{e^{-\theta C_i}}{\sum_{k=1}^N e^{-\theta C_k}} .
    \end{split}
\end{equation}

It is noticed that Equation \ref{equ:hi_2} is equal to the equilibrium point of the model (Equation \ref{equ:hi_1}), and thus it is proved that the equilibrium exists and is unique.

\subsection{Stability Analysis}

We now prove that $\vec f$ is asymptotically stable by using the following {\bf Lyapunov Theorem} \citep{Shafer1995Nonlinear,smith1984the,jin2007a}. 
For a dynamic system $\dot x= g(x)$, it is asymptotically stable at the stationary point $\bar x$, if there is a continuously differentiable scalar function $V(x)$ defined throughout the domain $\Psi$, such that 
\begin{itemize}
\vspace{-3 mm}
    \setlength{\itemsep}{0pt}
    \setlength{\parsep}{0pt}
    \setlength{\parskip}{0pt}

\item
$V(x)\geqslant 0$, $\forall x\in \Psi$;
\item 
$V(x)= 0$, if and only if $x$ is at the stationary point $\bar x$;
\item 
$V'(x)< 0$, if $x$ is not at the stationary point $\bar x$.
\end{itemize}

First, we construct a scalar function 
\begin{equation}
    V(\vec f)=z(\vec f)-\min z(\vec f)
\end{equation}

Then, it can be verified that
\begin{itemize}
\vspace{-3 mm}
    \setlength{\itemsep}{0pt}
    \setlength{\parsep}{0pt}
    \setlength{\parskip}{0pt}

\item
$V(\vec f)\geqslant 0$ in the whole domain;

\item 
$V(\vec f)= 0$ at the equilibrium point, since we have proven that the minimum value of $z(\vec f)$ can be reached at the equilibrium point;

\item 
$V'(\vec f)< 0$ is proved as follows.
\begin{equation}\label{equ:V'h_1}
    V'(\vec f) = \text{grad}V(\vec f)\dot{\vec f}
    = \sum_{i=1}^N[\theta C_i + \text{ln}(P_i f_i)+1]
    \left[ -P_i f_i + e^{-\theta C_i} \frac{\sum_{k=1}^N P_k f_k}{\sum_{k=1}^N e^{-\theta C_k}} \right]
\end{equation}
where $\text{grad}V(\vec f)$ is the gradient of $f_i$.
It is clear that
\begin{equation}
    \sum_{i=1}^N \left[ -P_i f_i + e^{-\theta C_i} \frac{\sum_{k=1}^N P_k f_k}{\sum_{k=1}^N e^{-\theta C_k}} \right] = 0.
\end{equation}
Thus, Equation \ref{equ:V'h_1} can be simplified to be the following equation.
\begin{equation}
    V'(\vec f) 
    = \sum_{i=1}^N[\theta C_i + \text{ln}(P_i f_i)]
    \left[ -P_i f_i + e^{-\theta C_i} \frac{\sum_{k=1}^N P_k f_k}{\sum_{k=1}^N e^{-\theta C_k}} \right].
\end{equation}
For any route $i$, if 
\begin{equation}
    -P_i f_i + e^{-\theta C_i} \frac{\sum_{k=1}^N P_k f_k}{\sum_{k=1}^N e^{-\theta C_k}} < 0,
\end{equation}
we have 
\begin{equation}\label{equ:V'h0_1}
\begin{split}
    [\theta C_i + \text{ln}(P_i f_i)] &
    \left[ -P_i f_i + e^{-\theta C_i} \frac{\sum_{k=1}^N P_k f_k}{\sum_{k=1}^N e^{-\theta C_k}} \right] \\ 
    & <
    \left[\theta C_i + \text{ln}\left(e^{-\theta C_i} \frac{\sum_{k=1}^N P_k f_k}{\sum_{k=1}^N e^{-\theta C_k}}\right)\right]
    \left[ -P_i f_i + e^{-\theta C_i} \frac{\sum_{k=1}^N P_k f_k}{\sum_{k=1}^N e^{-\theta C_k}} \right]   \\
    \Rightarrow \ \ \ \ [\theta C_i + \text{ln}(P_i f_i)] &
    \left[ -P_i f_i + e^{-\theta C_i} \frac{\sum_{k=1}^N P_k f_k}{\sum_{k=1}^N e^{-\theta C_k}} \right] \\ 
    & <
    \text{ln}\left(\frac{\sum_{k=1}^N P_k f_k}{\sum_{k=1}^N e^{-\theta C_k}}\right)
    \left[ -P_i f_i + e^{-\theta C_i} \frac{\sum_{k=1}^N P_k f_k}{\sum_{k=1}^N e^{-\theta C_k}} \right]. 
\end{split}     
\end{equation}
If 
\begin{equation}
    -P_i f_i + e^{-\theta C_i} \frac{\sum_{k=1}^N P_k f_k}{\sum_{k=1}^N e^{-\theta C_k}} > 0,
\end{equation}
we similarly have 
\begin{equation}\label{equ:V'h0_2}
\begin{split}
    [\theta C_i + \text{ln}(P_i f_i)] &
    \left[ -P_i f_i + e^{-\theta C_i} \frac{\sum_{k=1}^N P_k f_k}{\sum_{k=1}^N e^{-\theta C_k}} \right] \\ 
    & <
    \text{ln}\left(\frac{\sum_{k=1}^N P_k f_k}{\sum_{k=1}^N e^{-\theta C_k}}\right)
    \left[ -P_i f_i + e^{-\theta C_i} \frac{\sum_{k=1}^N P_k f_k}{\sum_{k=1}^N e^{-\theta C_k}} \right].
\end{split}     
\end{equation}
Combining Equations \ref{equ:V'h_1}, \ref{equ:V'h0_1} and \ref{equ:V'h0_2}, we finally prove $V'(f_i)<0$ as follows.
\begin{equation}
\begin{split}
    V'(\vec f) 
    & = \sum_{i=1}^N[\theta C_i + \text{ln}(P_i f_i)+1]
    \left[ -P_i f_i + e^{-\theta C_i} \frac{\sum_{k=1}^N P_k f_k}{\sum_{k=1}^N e^{-\theta C_k}} \right]\\
    & < \text{ln}\left(\frac{\sum_{k=1}^N P_k f_k}{\sum_{k=1}^N e^{-\theta C_k}}\right)\sum_{i=1}^N \left[ -P_i f_i + e^{-\theta C_i} \frac{\sum_{k=1}^N P_k f_k}{\sum_{k=1}^N e^{-\theta C_k}} \right]\\
    & =0.
\end{split}
\end{equation}

\end{itemize}

\section{Model Calibration and Validation}\label{sec:Calibration}

\subsection{Models for Testing}\label{sec:Testing}

The following three models are tested using the experimental data.
\begin{itemize}
\vspace{-3 mm}
    \setlength{\itemsep}{0pt}
    \setlength{\parsep}{0pt}
    \setlength{\parskip}{0pt}

\item
{\it Model A}: a full model. 
Without loss of generality, we set the form of $P_i$ and $C_i$ to be $P_i=1-\eta_i$ and $C_i=(1-\eta_i)c_i$, respectively.

\item 
{\it Model B}: a model with no consideration of the route-dependent attraction.
$\eta_i$ is equal to a constant $\eta$ and $C_i$ is independent to $\eta_i$.
For simplicity, we assume that the general cost on route $i$ is equal to the actual travel time, i.e., $C_i=c_i$. This is the exponential smoothing filter in \cite{Cantarella2016} with the Logit model as choice process and memory weighting parameter $\beta=1$. Here, we employ it as a baseline model.

\item 
{\it Model C}: an incomplete model that only takes into account the route-dependent inertia, i.e., $\eta_i\neq \eta_j$ ($i\neq j$), while $C_i$ is independent to $\eta_i$. 
We set $P_i=1-\eta_i$ and $C_i=c_i$. This model can illustrate if both route-dependent inertia and preference are necessary.
\end{itemize}
\vspace{-3 mm}

The above three models are all RUM-based models. 
The RBAP and PRC models are not tested here, since the RBAP model is an aggregate model without explicitly considering individual traveler's choice behavior and the PRC model largely deviates from the experimental observations as shown in Section \ref{Sec:observed_SR}.

The maximum likelihood estimation is used to estimate the models separately for each scenario, since subjects' choice behavior varies in different choice conditions. 
In the calibration process, the maximum likelihood estimation finds a group of parameters that can maximize log$L$, which is the natural logarithm of the likelihood function of the model.
For our models, it is as follows.
\begin{equation}
    \log L=\sum_{a} \sum_{t=1}^{N} \log \operatorname{Pr}(R_{a, t+1}=j_{a, t+1} \mid \theta, \eta_i, \vec{c_{a,t+1}}),
\end{equation}
where $j_{a, t+1}$ is the observed route choice of subject $a$ in round $t+1$ while $\sum_{a}$ is the summation over all subjects in the relevant scenarios; 
$\vec{c_{a,t+1}}$ is the cost combination experienced by subject $a$ in round $t+1$.

%

The following three performance indicators are introduced to assess the performance of the models.
\begin{itemize}
\vspace{-3 mm}
    \setlength{\itemsep}{0pt}
    \setlength{\parsep}{0pt}
    \setlength{\parskip}{0pt}
\item
Bayesian Information Criterion (BIC). BIC is a criterion for model selection among a finite set of models \citep{schwarz1978estimating,claeskens2008model}, and it is written as
\begin{equation}
\mathrm{BIC}=k\ln(n)-2\ln(\hat{L}),
\end{equation}
where $\hat L$ is the maximized value of the likelihood function of the model;
$n$ is the number of the observed data points;
$k$ is the number of the parameters estimated by the model.
It can be seen that BIC reflects the trade-off between the goodness of fit (likelihood) and simplicity of the model (number of parameters). The model with the lowest BIC is preferred.

\item 
Mean Absolute Percentage Error (MAPE) of the switching rate:
\begin{equation}
    \mathrm{MAPE}_p = \frac{1}{|\mathbf{C}|N^2}  \sum_{\vec{c}\in \mathbf{C}} \sum_{i=1}^N  \sum_{j=1}^N  \frac{| \tilde p_{ij}(\vec{c}) - \bar{p}_{ij}(\vec{c}) |}{\bar{p}_{ij}(\vec{c})},
\end{equation}
where $|\mathbf{C}|$ is the size of the set of cost combinations in a scenario. Only cost combinations occurring more than a certain threshold $\theta$ (here we choose $\theta=8$) are considered to avoid random noises.

\item 
MAPE of the equilibrium flow:
\begin{equation}
    \mathrm{MAPE}_f = \frac{1}{N}  \sum_{i=1}^N \frac{| \tilde f_i - \bar f_i |}{\bar f_i},
\end{equation}
where $\bar f_i$ and $\tilde f_i$ is the mean experimental flow and the model-estimated equilibrium flow of route $i$, respectively. 
As the route flow keeps fluctuating in the experiments, we take the mean flow volume (presented in Table \ref{tab:Statistics_Time}) as an approximation of the equilibrium flow.
\end{itemize}

The above three indicators evaluate the performances of models from two different perspectives. BIC is to measure relative performances of different models, while MAPEs could reflect to what extent the models can predict the data.

Table \ref{tab:calibration} presents the calibration results and the performances of Models A, B, and C. 
It can be seen that (i) Model A achieves the lowest MAPE$_p$ and MAPE$_f$ in all eight scenarios, and it also results in the best BIC, indicating that the additional parameters (comparing to Model B) are indispensable.
(ii) Model C performs satisfactorily only in the two-route scenarios; 
(iii) comparing to other models, Model B is inaccurate in all scenarios except the symmetric Scenario 1.
The performance of the models will be explicitly analyzed in the following two subsections.

\begin{table}[!htbp]\centering\footnotesize
\setlength{\tabcolsep}{3.5mm}{
\caption{Calibration results of Models A, B and C in the eight scenarios.}\label{tab:calibration}
\begin{tabular}{ccllllllll}
\toprule
Scenario & Model & $\theta$ & $\eta_1$ & $\eta_2$ & $\eta_3$ & $\text{MAPE}_p$ & $\text{MAPE}_f$ & BIC \\
\midrule
1 &  A & 0.0683 & 0.359 & 0.351 & --- & 0.0896 & 0.0001 & 5647.002\\ 
 &  B & 0.0439 & 0.355 & 0.355 & --- & 0.0883 & 0.00563 & 5581.657\\
 &  C & 0.0441 & 0.363 & 0.346 & --- & 0.0899 & 0.00755 & 5647.048
 \vspace{2mm}\\
2 &  A & 0.0525 & 0.555 & 0.403 & --- & 0.135 & 0.004 & 4845.526\\
 &  B & 0.0349 & 0.532 & 0.532 & --- & 0.230 & 0.145 & 4955.487\\
 &  C & 0.0305 & 0.648 & 0.294 & --- & 0.143 & 0.0082 & 4853.818
 \vspace{2mm}\\
3 &  A & 0.0665 & 0.518 & 0.307 & --- & 0.112 & 0.0058 & 4854.039\\
 &  B & 0.0493 & 0.471 & 0.471 & --- & 0.277 & 0.211 & 4996.124\\
 &  C & 0.0393 & 0.603 & 0.192 & --- & 0.118 & 0.0034 & 4855.359
 \vspace{2mm}\\
4 &  A & 0.0204 & 0.443 & 0.197 & --- & 0.0389 & 0.059 & 5980.376\\
 &  B & 0.0208 & 0.377 & 0.377 & --- & 0.245 & 0.181 & 6139.933\\
 &  C & 0.0133 & 0.517 & 0.094 & --- & 0.0508 & 0.0072 & 5983.256
 \vspace{2mm}\\
 5 &  A & 0.0421 & 0.468 & 0.235 & --- & 0.0571 & 0.007 & 6975.715\\
 &  B & 0.00415 & 0.405 & 0.405 & --- & 0.223 & 0.223 & 7145.333\\
 &  C & 0.0271 & 0.537 & 0.140 & --- & 0.0583 & 0.0110 & 6977.567
 \vspace{2mm}\\
6 & A & 0.0168 & 0.480 & 0.315 & 0.125 & 0.197 & 0.0419 & 13936.454\\
 & B & 0.00302 & 0.362 & 0.362 & 0.362 & 0.337 & 0.337 & 14204.863\\
 & C & 0.0158 & 0.511 & 0.291 & 0.0867 & 0.250 & 0.134 & 14026.736
 \vspace{2mm}\\
7 & A & 0.0266& 0.439 & 0.218 & 0.0976 & 0.185 & 0.0504 & 14577.203\\
 & B & 0.00966 & 0.303 & 0.303 & 0.303 & 0.323 & 0.321 & 14842.366\\
 & C & 0.00633 & 0.467 & 0.186 & 0.0775 & 0.237 & 0.125 & 14652.433
 \vspace{2mm}\\
8 & A & 0.00875& 0.516 & 0.319 & 0.116 & 0.288 & 0.0784 & 18192.628\\
 & B & 0.00237 & 0.383 & 0.383 & 0.383 & 0.419 & 0.3354 & 18615.532\\
 & C & 0.00117 & 0.538 & 0.296 & 0.0863 & 0.338 & 0.0824 & 18497.270\\
\bottomrule
\end{tabular}}
\end{table}

\subsection{Model-based Estimation Results}

This subsection separately presents and analyzes the estimation results of Models A, B and C in the two-route and three-route scenarios. 

\subsubsection{Two-Route Scenarios}

From Table \ref{tab:calibration}, it can be seen that Models A and C have similar performance indicated by all three performance indicators. In contrast, Model B, which ignores the route-dependent attraction, performs the worst in all scenarios.
The results indicate that subjects' choice behavior in two-route scenarios can be measured by using only the route-dependent inertia. 
The reason is that when facing a binary choice (i.e. if switching or not) in a two-route scenario, the co-effects of route-dependent inertia and preference (in Model A) could be substituted by larger route-dependent inertia (in Model C). 



Moreover, it is noticed that Model B also has good performance in Scenario 1.
It is probably because the symmetry routes in Scenario 1 have the same attraction to the subjects, implying the importance of the route-dependent attraction.

\begin{figure}[!htbp]
    \centering
    \includegraphics[width=6.5in]{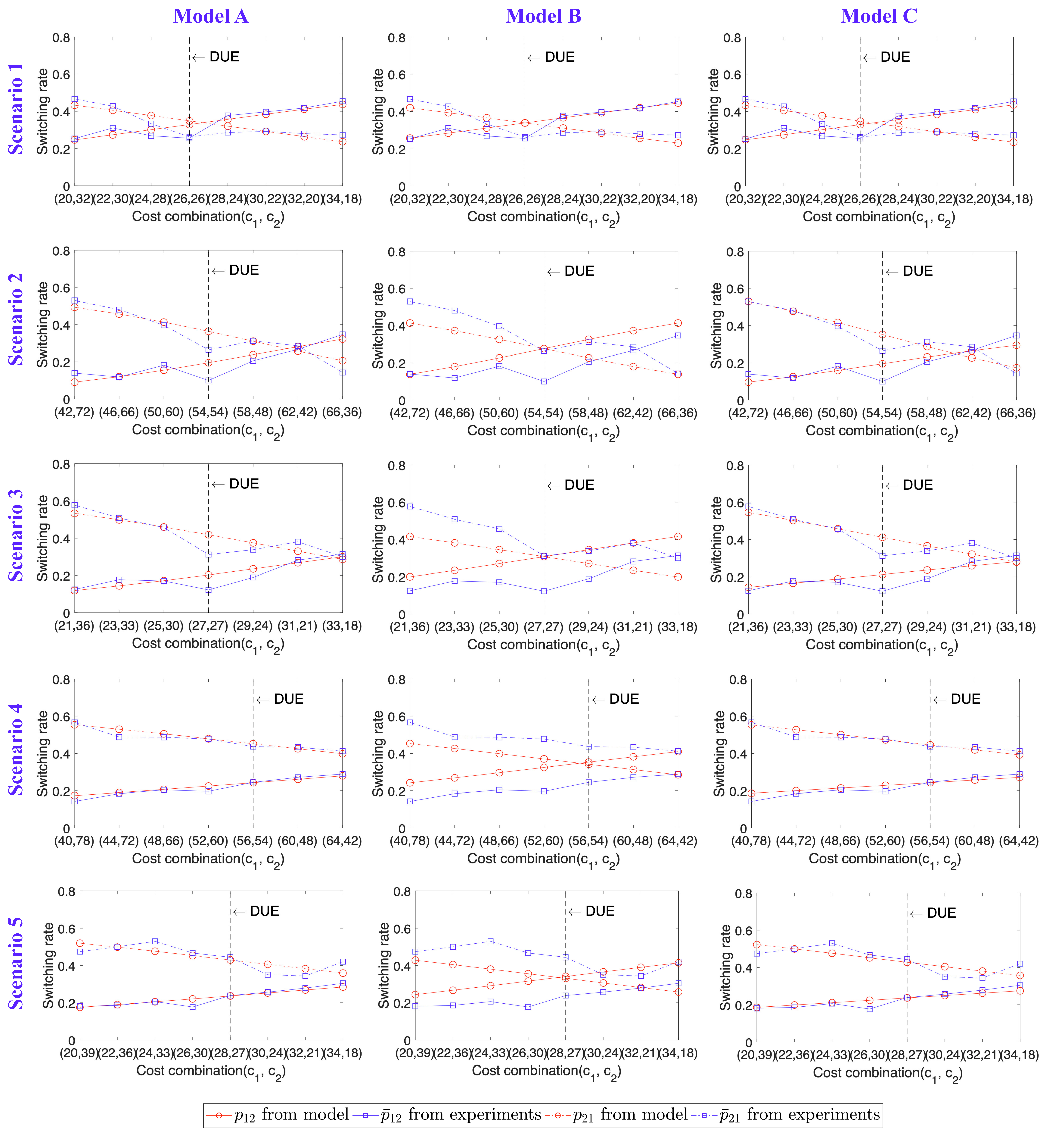}
    \caption{Detailed comparisons between average experimental switching rates and theoretical switching rates in all the two-route scenarios (Scenarios 1-5).}
    \label{fig:TwoRoute_Switch}
\end{figure}

To see more details, Figure \ref{fig:TwoRoute_Switch} compares the model-based estimation of the switching rates with the experimental results by presenting the switching rates of both routes at every cost combination. 
It can be found that the experimental results can be well reproduced by Models A and C.
However, for Model B, the estimated switching rates of the two routes (i.e., $p_{12}$ and $p_{21}$) overlap with each other at the DUE point, and thus a large deviation from the experimental results occurs. 
This observation is resulted from the absence of the route-dependent attraction in Model B. 


\subsubsection{Three-Route Scenarios}\label{sec:three-route}

Table \ref{tab:calibration} shows that Model B is the worst one among the three models in the three-route scenarios (Scenarios 6-8), which is consistent with the results in two-route scenarios.
However, different from the two-route scenarios, Model A outperforms Model C in the three-route scenarios.

Figures \ref{fig:ThreeRoute_Switch} and \ref{fig:ThreeRoute_MAPE_Detail} present the detailed switching rates in the three-route scenario, further confirming that Model A performs the best in predicting the switching rates. It is interesting to observe that Models A and C perform similarly when predicting $p_{ii}$ (i.e. $p_{11}$, $p_{22}$ and $p_{33}$). However, Model A achieves much lower errors when describing other switching rates.

The above observation indicates that both route-dependent inertia and preference have significant impacts on the DTD route choices.
We conjecture the reason that Model A (i.e., the full model) outperforms Model C (i.e., the model only considering route-dependent inertia) in the three-route scenarios as follows.
When facing simple binary-choice problem in the two-route scenarios (i.e., whether or not to stay on the current route), subjects make route choices probably more dependent on (the route-dependent) inertia. In contrast, when subjects have more than one options in the three-route scenarios, the route-dependent preference would play a more important role.


\begin{figure}[!htbp]
    \centering
    \subfigure[Scenario 6]{
    \includegraphics[width=2.6in]{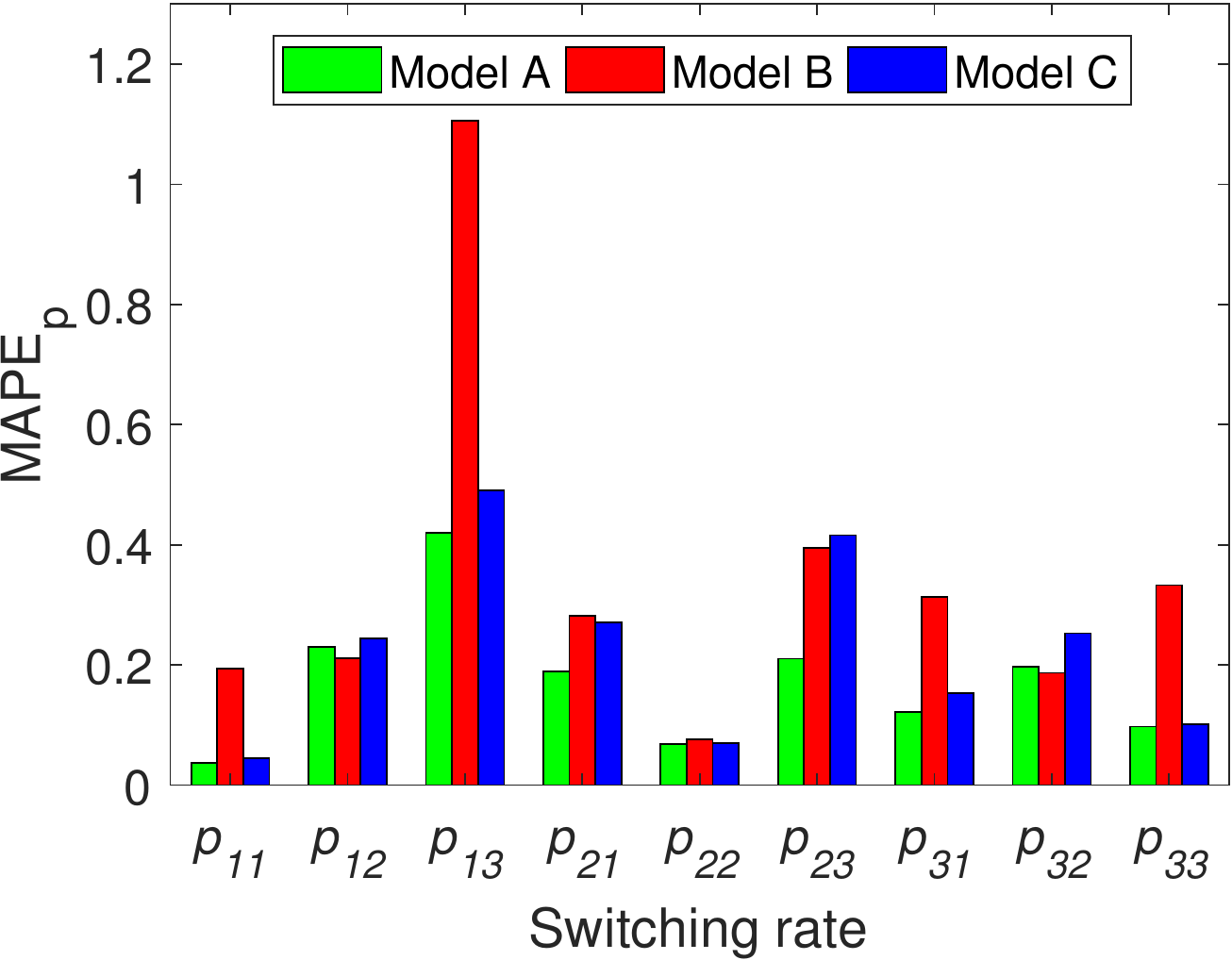}}
    \hspace{10mm}
    \subfigure[Scenario 7]{
    \includegraphics[width=2.6in]{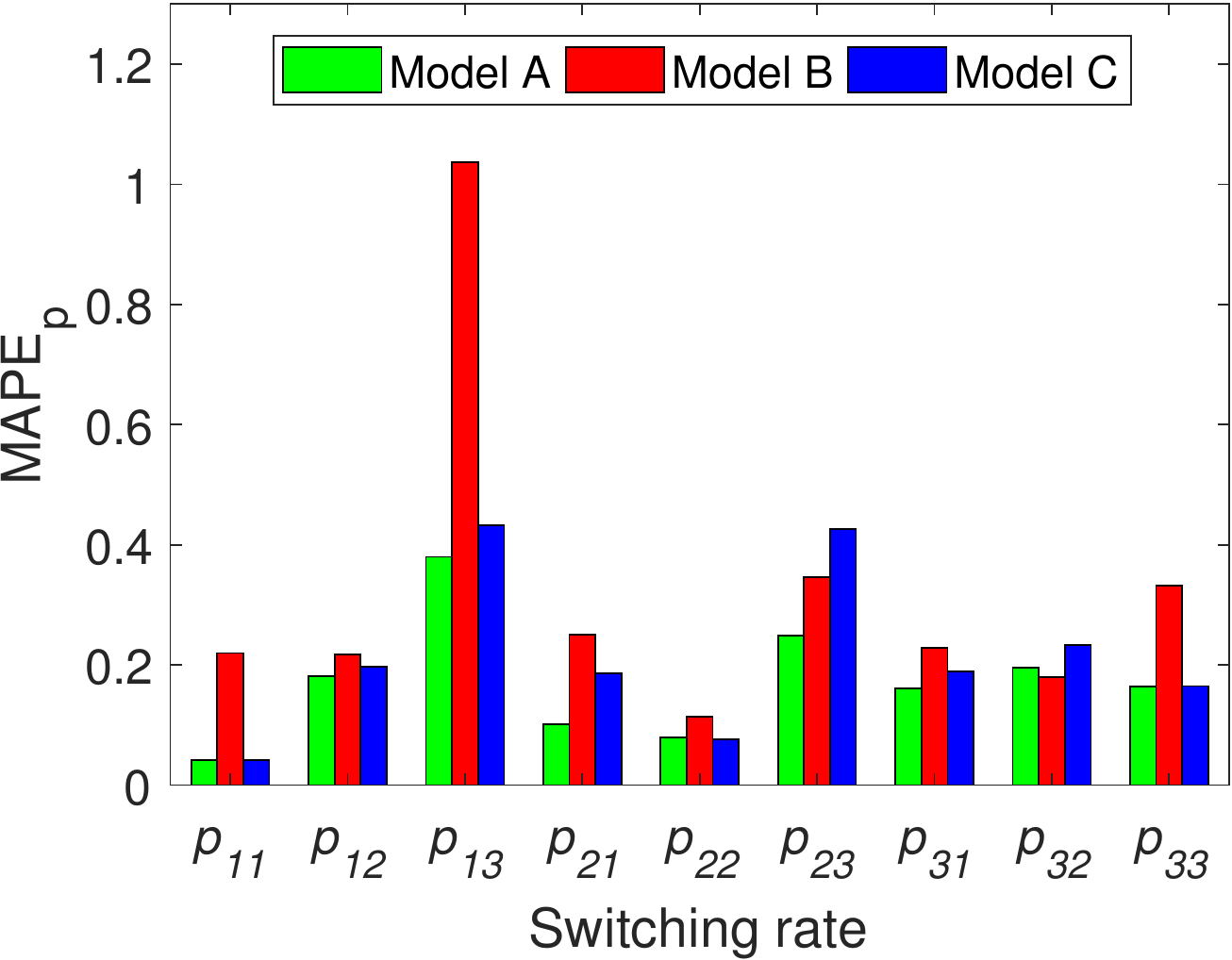}}
    \hspace{10mm}
    \subfigure[Scenario 8]{
    \includegraphics[width=2.6in]{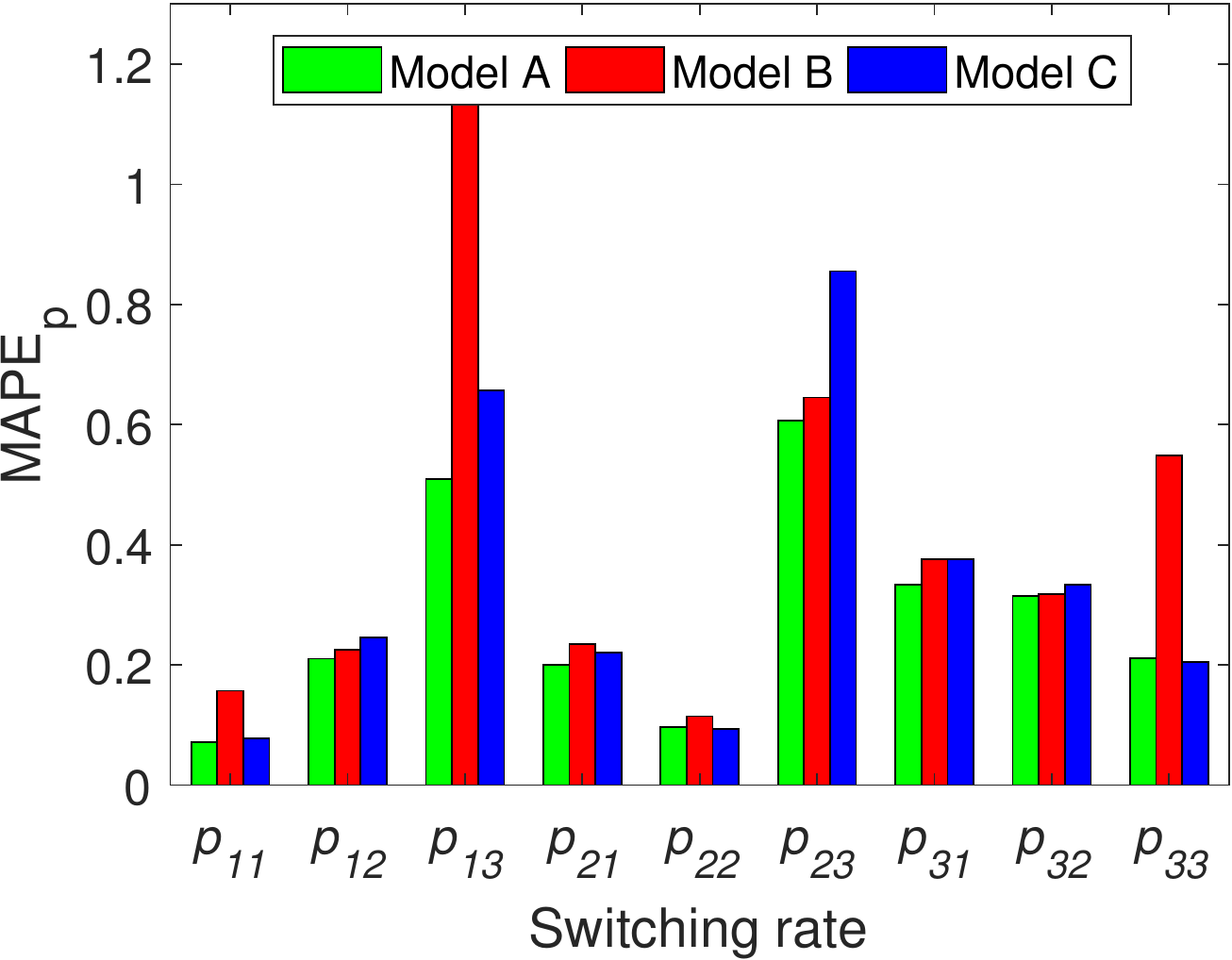}}
    \caption{Detailed comparisons of route-wise switching rates estimated by Models A, B, and C in the three-route scenarios (Scenarios 6-8).}
    \label{fig:ThreeRoute_Switch}
\end{figure}

\begin{landscape}
    \begin{figure}[!htbp]\ContinuedFloat*
        \centering
        \includegraphics[width=9in]{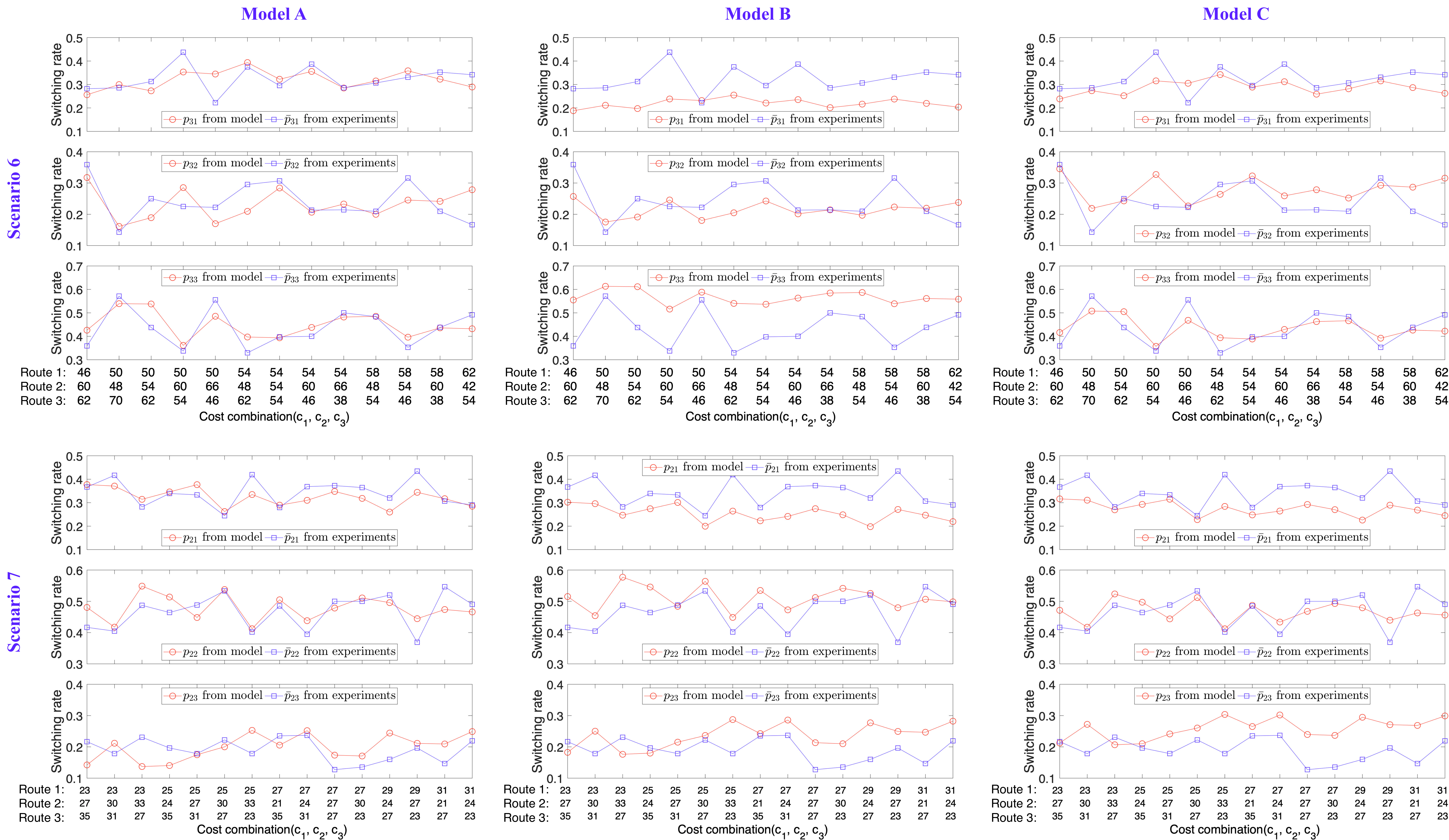}
        \caption{Detailed comparison between the average experimental switching rates and theoretical switching rates in the three-route scenarios (Scenarios 6-8). }
        \label{fig:ThreeRoute_MAPE_Detail}
    \end{figure}
\end{landscape}

 \begin{figure}[!hbp]\ContinuedFloat
    \centering
    \includegraphics[width=5.2in]{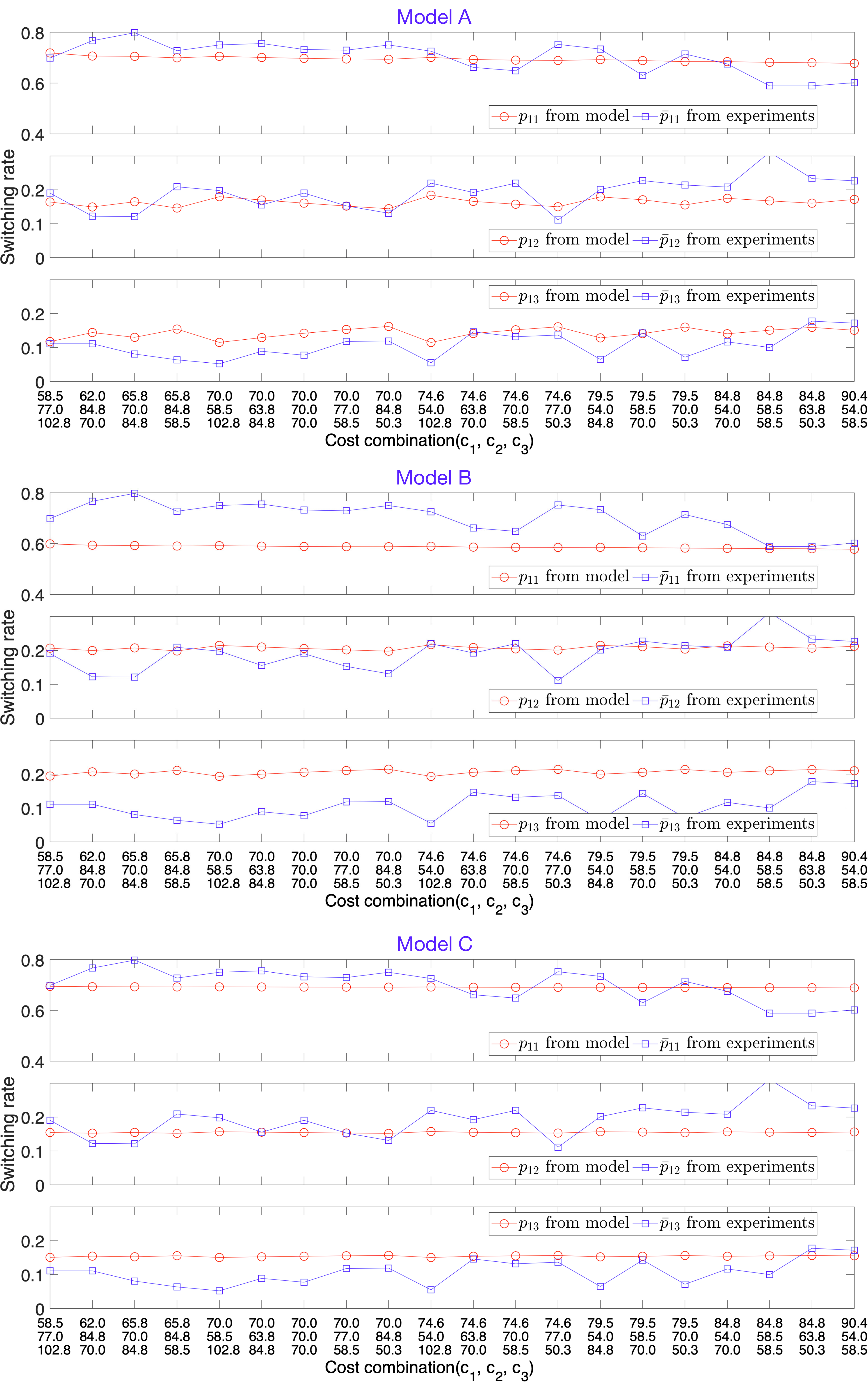}
    \caption{Continued.}
\end{figure}

In summary, the proposed model (i.e., Model A) can satisfactorily reproduce the experimental results, including the average switching rate and the equilibrium flow.
It is well shown that the proposed assumption of route-dependent attractions, including route-dependent inertia and preference, is critical for the performance of the model and for describing the DTD route choice dynamics.

\subsection{Model Validation}

The validation of the proposed model is further enhanced from two different aspects. 
One is to conduct an out-of-sample test that is different from the calibration process, and the other is to validate the model by using the experiment data reported in the existing studies.

\subsubsection{Out-of-sample Test}
The out-of-sample test is a widely used method of testing the predictive ability of a model.  
To do this, the dataset is randomly divided into two parts, i.e., 75\% and 25\%, respectively.
The first part is used to estimate the parameters and the second part is used to validate the model, i.e. to calculate the out-of-sample MAPE and Log-likelihood values as performance indices of the model. 
It can be seen from the results presented in Table \ref{tab:out-of-sample} that (1) the MAPEs increased with limited magnitudes; (2) among the three models, Model A performs the best in terms of both MAPE and Log-likelihood; (3) the calibrated parameters also change within a limited range. The above observations well demonstrate the robustness of the proposed model.

\begin{table}[!htbp]\centering\footnotesize
\setlength{\tabcolsep}{5mm}{
\caption{The out-of-sample test results of Models A, B, and C in the eight scenarios.}\label{tab:out-of-sample}
\begin{tabular}{ccllllll}
\toprule
Scenario & Model & $\theta$ & $\eta_1$ & $\eta_2$ & $\eta_3$ &$\text{MAPE}_p$ & LL\\
\midrule
 &  A & 0.0766 & 0.375 & 0.357 & --- & 0.105 & -696.451 \\
1 &  B & 0.0482 & 0.367 & 0.367 & --- & 0.112 & -695.559 \\
 &  C & 0.0484 & 0.3384 & 0.348 & --- & 0.0990  & -696.240 \vspace{2mm}\\
 &  A & 0.0499 & 0.510 & 0.367 & --- & 0.153 & -547.720  \\
2 &  B & 0.0347 & 0.485 & 0.485 & --- & 0.287 & -577.417 \\
 &  C & 0.0310 & 0.603 & 0.262 & --- & 0.156 & -546.614 \vspace{2mm}\\
 &  A & 0.0588 & 0.513 & 0.299 & --- & 0.165 & -507.918 \\
3 &  B & 0.0441 & 0.459 & 0.459 & --- & 0.318 & -535.869 \\
 &  C & 0.0336 & 0.590 & 0.191& --- & 0.169 & -507.673  \vspace{2mm}\\
 & A & 0.0195 & 0.460 & 0.190 & --- & 0.0991 & -977.175 \\
4 & B & 0.0208 & 0.390 & 0.390 & --- & 0.285 & -1004.762 \\
 & C & 0.0128 & 0.808 & 0.535& --- & 0.107 & -977.804  \vspace{2mm}\\
 & A & 0.0287 & 0.474 & 0.241 & --- & 0.0988 & -1141.048 \\
5 & B & 0.0304 & 0.402 & 0.402 & --- & 0.281 & -1213.455 \\
 & C & 0.0197 & 0.522 & 0.176 & --- & 0.112 & -1146.382 \vspace{2mm}\\
 & A & 0.0184 & 0.474 & 0.334 & 0.142 & 0.271 & -1698.518 \\
6 & B & 0.00394 & 0.365 & 0.365 & 0.365 & 0.450 & -1766.249 \\
 & C & 0.00255 & 0.501 & 0.317 & 0.101 & 0.323 & -1707.401  \vspace{2mm}\\
 & A & 0.0302 & 0.419 & 0.249 & 0.0830 & 0.265 & -1904.96 \\
7 & B & 0.00636 & 0.298 & 0.298 & 0.298 & 0.431 & -2049.229 \\
 & C & 0.00327 & 0.439 & 0.229 & 0.0562 & 0.310 & -1940.712 \vspace{2mm}\\ 
 & A & 0.00807 & 0.543 & 0.314 & 0.138 & 0.396 & -2948.421 \\
8 & B & 0.00224 & 0.404 & 0.404 & 0.404 & 0.534 & -3148.511 \\
 & C & 0.00059 & 0.566 & 0.283 & 0.122 & 0.449 & -2987.147 \\ 
\bottomrule
\end{tabular}
}

\end{table}

\subsubsection{Testing with Other Data Sources}

\begin{figure}[!hbp]
    \centering
    \subfigure[Model A]{
    \includegraphics[width=3in]{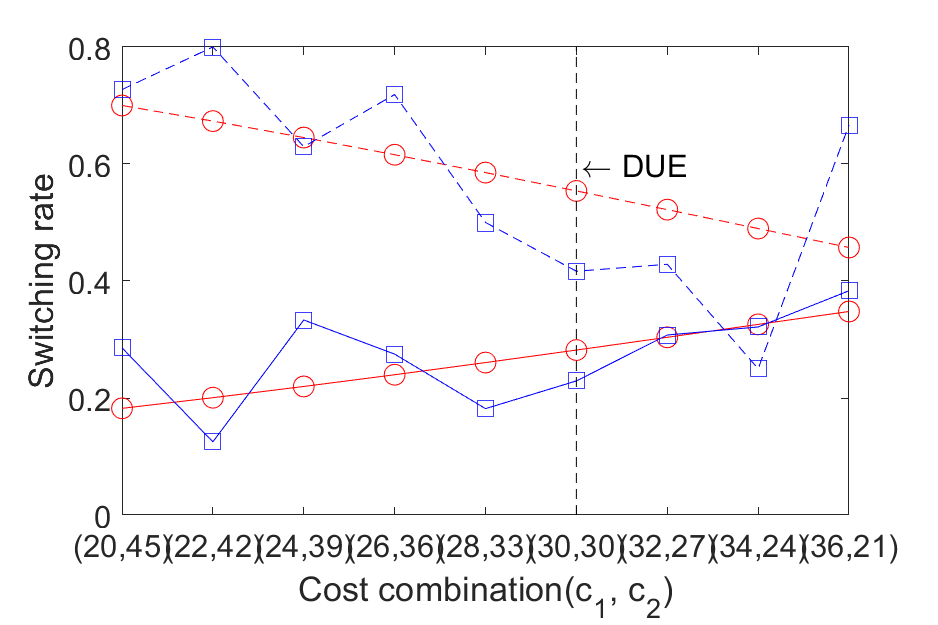}}
    \subfigure[Model B]{
    \includegraphics[width=3in]{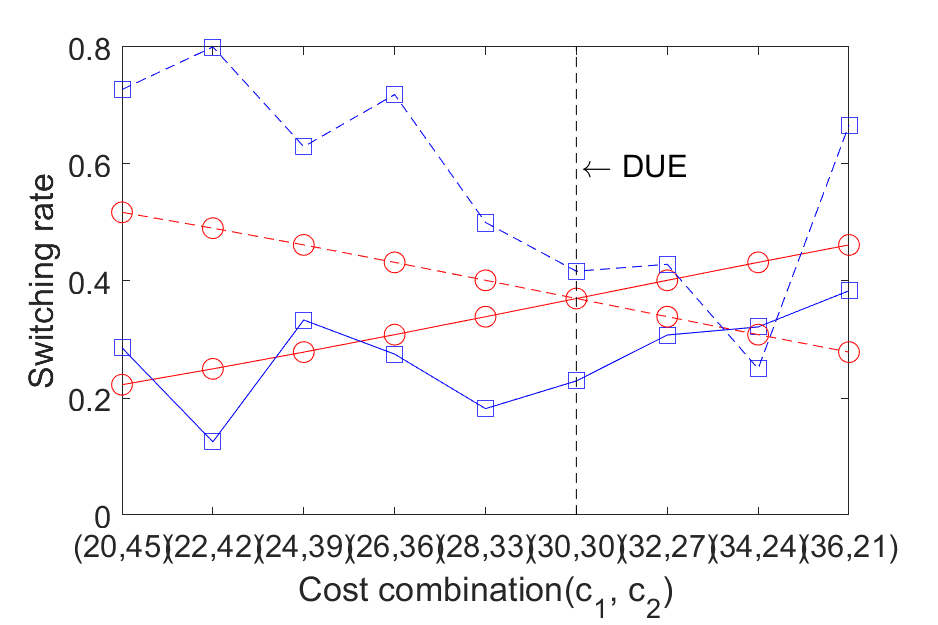}}
    \subfigure[Model C]{
    \includegraphics[width=3in]{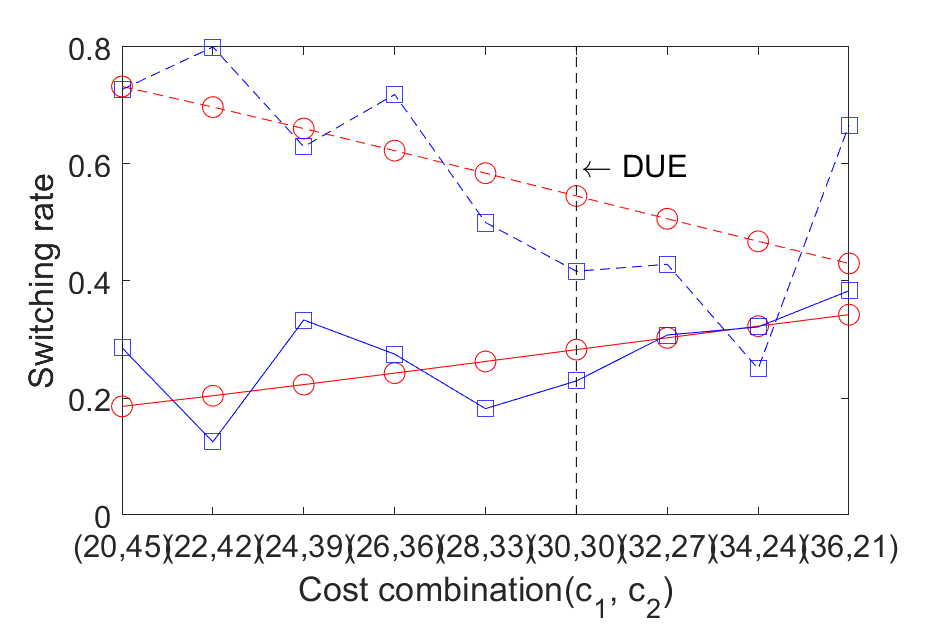}}    
    \subfigure{
    \includegraphics[width=3in]{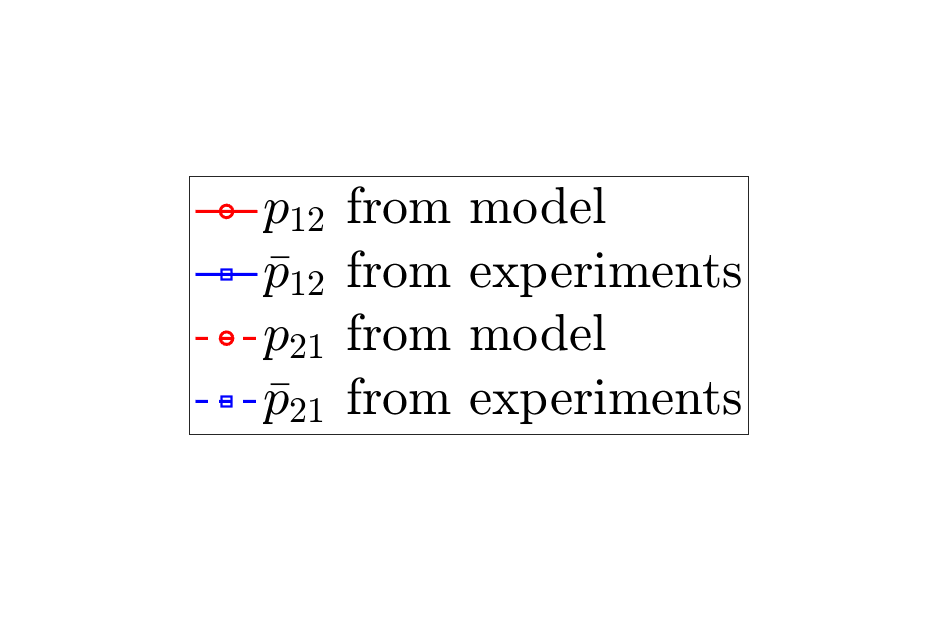}}
   
    \caption{Result comparisons of Models A, B, and C between the model estimation and the observed data in \cite{Zhao2016c}.  }
    \label{fig:Zhao}
\end{figure}

To show the generalization of the proposed model, we test it by using another dataset in \cite{Zhao2016c}, where 18 subjects were invited into a laboratory to repeatedly make decisions for 54 rounds in a two-route network. 
The individual-level route choice data is presented in \cite{Zhao2016c} so that we could apply the method in Section \ref{sec:Testing} to calibrate and test the proposed model.
The calibration results are presented in Table \ref{tab:Zhao}, and the estimation results of the proposed model are compared with the experiment data in Figure \ref{fig:Zhao}. 
It can be seen from Figure \ref{fig:Zhao} that the results are similar to those based on our experiments, i.e., both Models A and C have good performance in reproducing the DTD route choices, while Model B is the worst. 
It further confirms the importance of introducing route-dependent attractions.

\begin{table}[!htbp]\centering\footnotesize
\setlength{\tabcolsep}{6mm}{
\caption{Calibration results using the dataset in \cite{Zhao2016c}.}}\label{tab:Zhao}
\begin{tabular}{ccccccc}
\toprule
Model & $\theta$ & $\eta_1$ & $\eta_2$ & $\text{MAPE}_p$ & $\text{MAPE}_f$ & BIC\\
\midrule
 A & 0.0317 & 0.340 & 0.0327 & 0.195 & 0.04   &  737.451\\
 B & 0.0337 & 0.260 & 0.261 & 0.323 & 0.189  &  752.291\\
 C & 0.0286 & 0.435 & 0.0119 & 0.186 & 0.039 &  736.723
\\
\bottomrule
\end{tabular}
\end{table}

\section{Illustrative Numerical Examples}\label{sec:Example}

Equation \ref{equ:hi_1} shows that SUE would be a special case of the proposed model when all route-dependent attraction coefficients equal zero. 
However, if there are non-zero attraction coefficients, the equilibrium state of the proposed model would deviate from SUE.
The route-dependent attraction plays two roles: (1) the route-dependent inertia hinders changing routes; (2) the route-dependent preference weights the utility of different routes. 
Combining the two effects, a route with a higher attraction coefficient implies that it would attract more travelers during the evolution process and has a higher flow at the equilibrium state.

This section is to show the above characteristics of the proposed model by using an illustrative numerical example. 
Instead of using a complex network, we employ a simple two-route network for better illustration since the primary goal of the example is to compare the proposed model with the classic deterministic and stochastic models. 
Two scenarios including a symmetrical case and an asymmetrical case are carried out, and the cost functions are presented in Table \ref{tab:numericalexamples}.

\begin{table}[!htbp]\center\footnotesize
\setlength{\tabcolsep}{17mm}{
\caption{Parameters setting in the numerical examples.}\label{tab:numericalexamples}
\begin{tabular}{ccc}
\toprule
{\footnotesize Scenario} & {\footnotesize Cost function} & {\footnotesize DUE$^{*}$} \\
\midrule
 1 & \tabincell{c}{ $c_{1}=20(1+0.15({f_{1}}/{0.5})^{2})$ \\ $c_{2}=20(1+0.15({f_{2}}/{0.5})^{2})$ }  &  0.5\ ,\ 0.5\ \vspace{2mm}\\
 2 & \tabincell{c}{ $c_{1}=20(1+0.15({f_{1}}/{0.6})^{2})$ \\ $c_{2}=20(1+0.15({f_{2}}/{0.4})^{2})$ }  &  0.6\ ,\ 0.4\ \\
\bottomrule
\end{tabular}}
\begin{tablenotes}
\item[1]$^{*}$ DUE flow assignment; The unit is ``traveler".
 \end{tablenotes}
\end{table}

\begin{figure}[!htbp]
    \centering
    \subfigure[Scenario 1]{
    \includegraphics[width=3in]{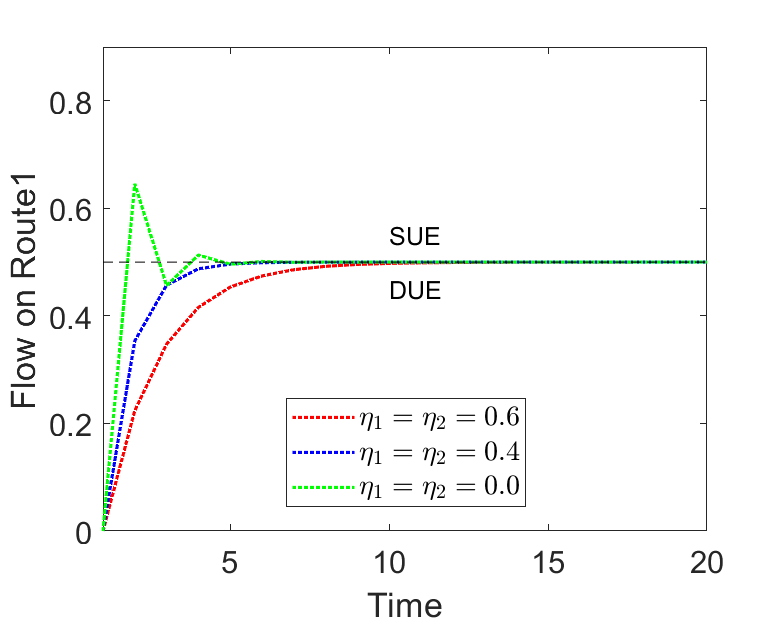}}
    \subfigure[Scenario 2]{
    \includegraphics[width=3in]{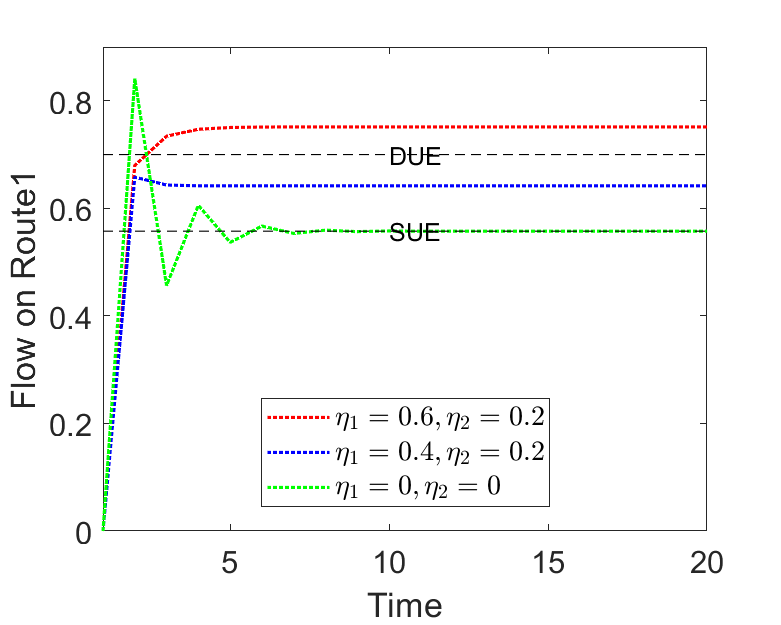}}
    \caption{Results of the numerical tests. The traffic flow on the route 1 is taken as an example.}
    \label{fig:Num}
\end{figure}

Figure \ref{fig:Num} presents the numerical results, which are consistent with the analysis at the beginning of this section.
Specifically, in the symmetrical case, the two routes have the same travel time and flow, indicating the equilibrium flow assignment is identical to both DUE and SUE. 
However, the system spends more time on converging before reaching the equilibrium state when the route-dependent attraction coefficient is large.
The slow convergence is caused by the larger route-dependent inertia that makes travelers less likely to change routes.

In the asymmetrical case, the equilibrium flow assignment of the proposed model is the same with SUE when all attraction coefficients are zero, as predicted.
However, it is obviously different from DUE and SUE when the coefficients are not zero. 
Moreover, the equilibrium flow assignment of the proposed model varies with different route-dependent attraction coefficients. The results agree with our conjecture, i.e., the higher attraction coefficient turns out, the higher flow volume at the equilibrium state.

\section{Discussion}\label{sec:Discussion}
\subsection{Reproduction of Flow Oscillations}
We discuss the possible reasons behind the non-convergence tendency of network flow evolution as follows. One reason may lie in that the route-choice games in our experiments are essentially non-cooperative \textit{N}-person games with multiple equilibria. The number of multiple equilibria depends on the architecture of the network and the number of network users, which may be as large as thousands or occasionally in millions. Taking Scenario 2 in the paper as an example, although all the equilibria share the same flow distribution (5 players on the upper and 11 subjects on the bottom route), the possible permutations of 5 players among a group of 16 people vary across different equilibria. For another example, \citet{Mak2015} examined a very ``complex" network (for experimental research) with 12 links and 8 routes and 18 players participated in each session of the experiments. There are over 3.3 billion pure-strategy equilibria in their experiments. Therefore, as commented by \citet{Rapoport2019book} in a book chapter, ``\textit{if users behaved `as if' they possessed common knowledge of full rationality, it would still have been virtually impossible for them to coordinate to attain any specific equilibrium}".

To fill the gap, this subsection evaluates the proposed model from the dynamical perspective by using network flow oscillations over time.

Suppose that all travelers are homogeneous, and the switch rate $p^t_{ij}$  can be viewed as the probability that an individual traveler switches to route $j$ from route $i$. Thus, the switch flow from route $i$ to route $j$ (i.e., $\Delta f^t_{ij}$ in Equation \ref{equ:Delta_h}) is the sum of $f^t_i$ independent
Bernoulli trials with the probability of success $p^t_{ij}$. Therefore, $\Delta f^t_{ij}$ is a stochastic variable that obeys a Binomial distribution, i.e.,

\begin{equation}\label{equ:stochastic model}
    \Delta f_{ij}^t \sim \text{Binomial} (f_{i}^t, {p}_{ij}^t).
\end{equation}

\begin{figure}[!htbp]
    \centering
    \centering
    \subfigure[Evolution process]{
    \includegraphics[width=3in]{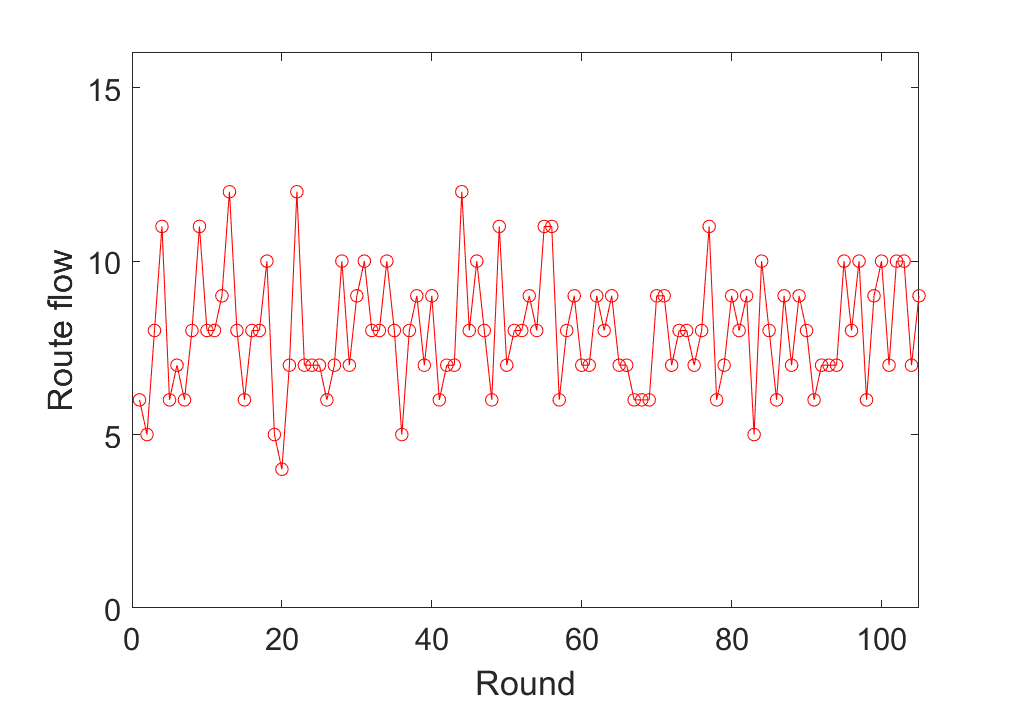}}
    \subfigure[Route flow distribution]{
    \includegraphics[width=3in]{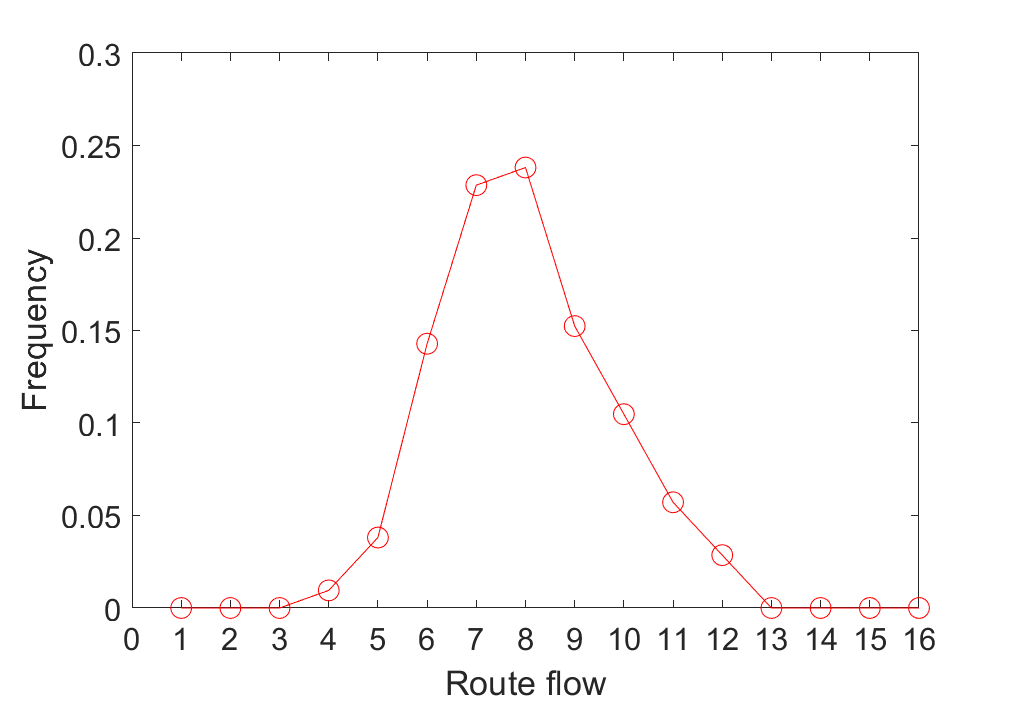}}
    \caption{An example of route flow evolution and distribution (Session 3 in Scenario 2).}
    \label{fig:distribution}
\end{figure}

According to Equation \ref{equ:Delta_h}, the flow on route $i$ at time $t+1$ (i.e., $f^{t+1}_i$ ) is the sum of Bernoulli distributed variable $\Delta f^t_{ki}$. Therefore, $f^{t+1}_i$ follows the Possion bi-nominal distribution, which is determined by and only by the network flow on all routes at time $t$
(denoted by $\vec{F}^{t}$). The fact that $\vec{F}^{t+1}$ depends on and only on $\vec{F}^{t}$ indicates that the evolution of the network flow is a Markov process. Therefore, it is not surprise that the route flow keeps fluctuating all the time. In the given Markov process, the transition probabilities between any two states are non-zero. Thus, it is an irreducible and aperiodic Markov chain with finite state space and a unique equilibrium distribution exists for the Markov process \citep{serfozo2009basics}.
Figure \ref{fig:distribution} illustrates an example of route flow evolution and its distribution of Session 3 in Scenario 2.

\begin{figure}[!htbp]
    \centering
    \subfigure[Scenario 1]{
    \includegraphics[width=3in]{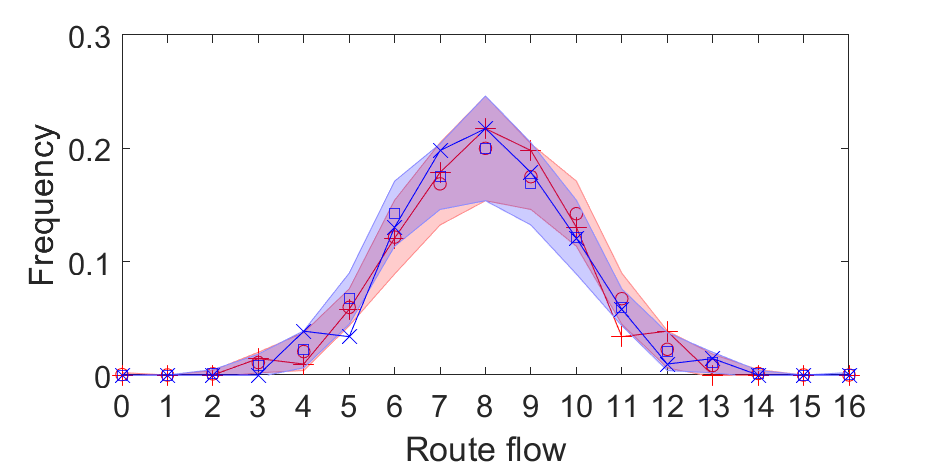}}
    \subfigure[Scenario 2]{
    \includegraphics[width=3in]{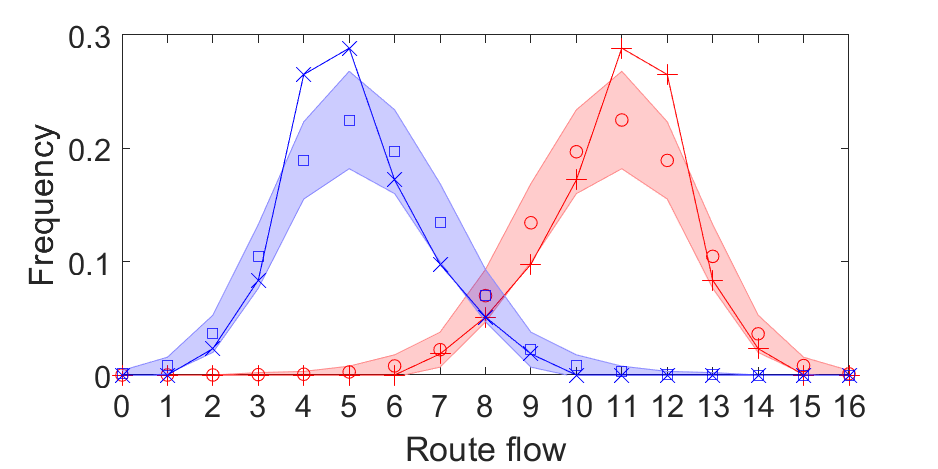}}
    %
    \subfigure[Scenario 3]{
    \includegraphics[width=3in]{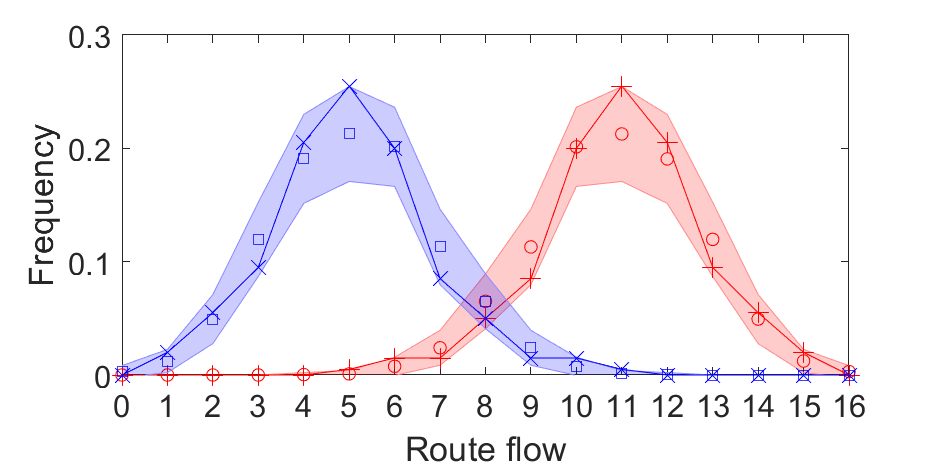}}
    \subfigure[Scenario 4]{
    \includegraphics[width=3in]{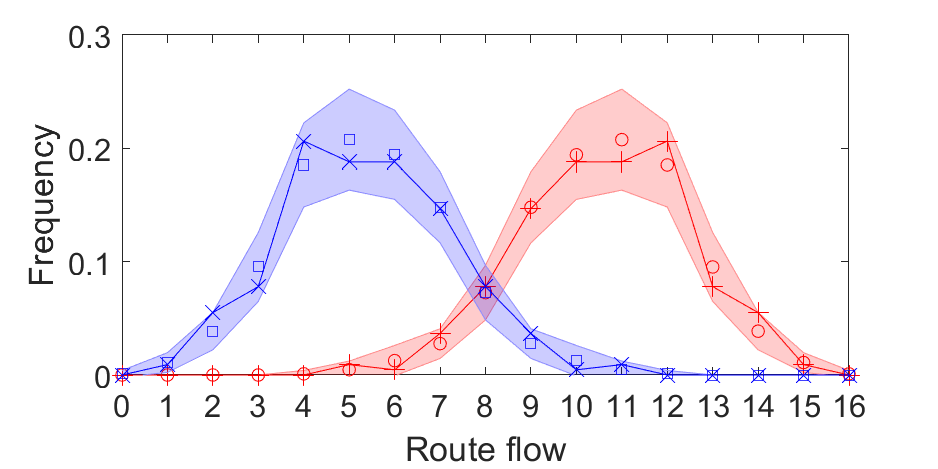}}
    %
    \subfigure[Scenario 5]{
    \includegraphics[width=3in]{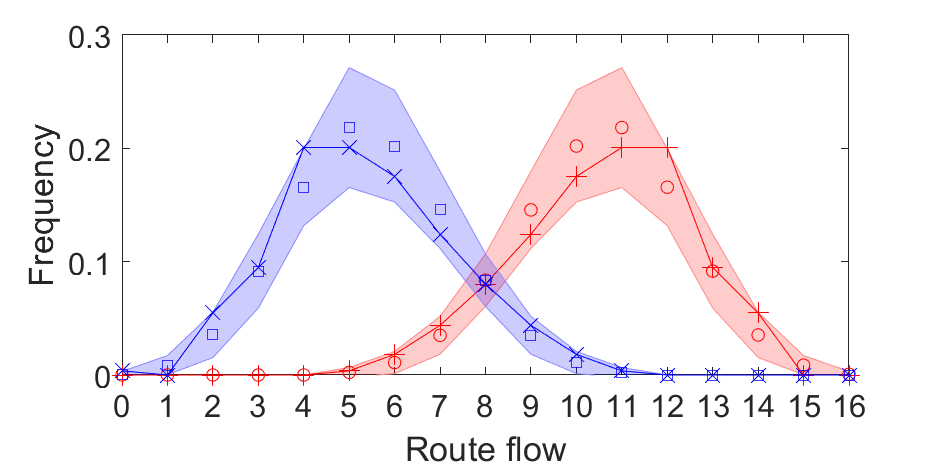}}    
    \subfigure[Scenario 6]{
    \includegraphics[width=3in]{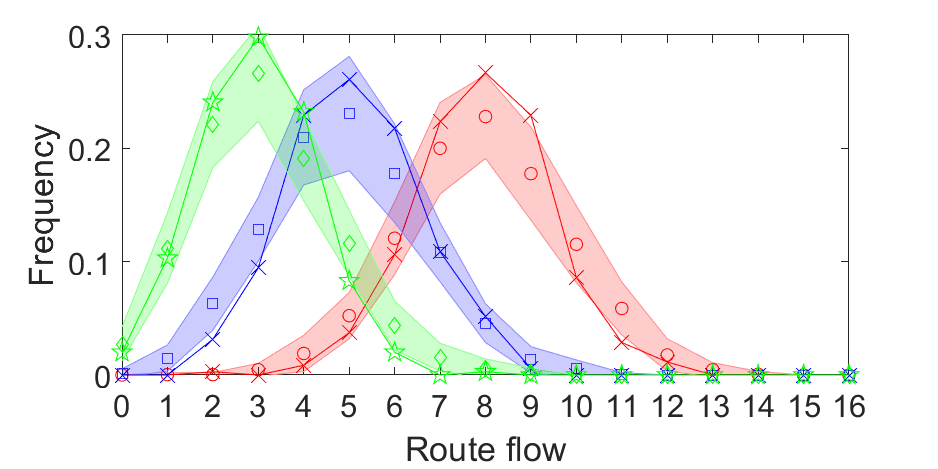}}
    %
    \subfigure[Scenario 7]{
    \includegraphics[width=3in]{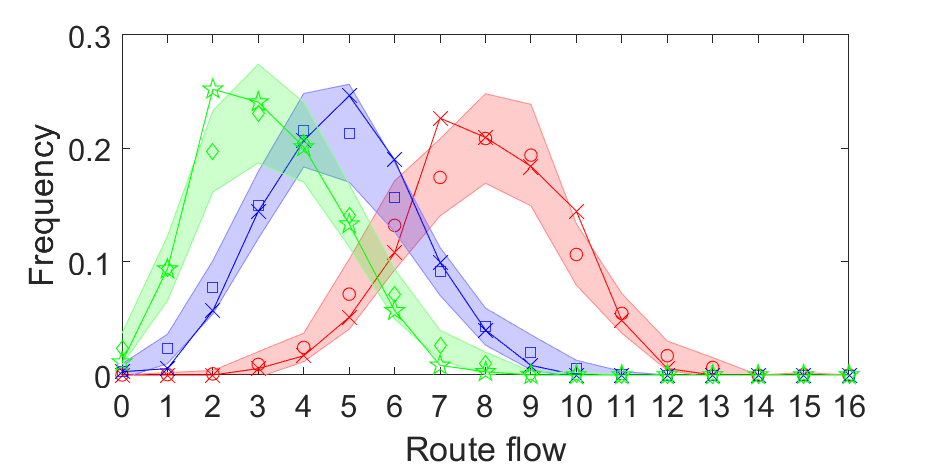}} 
    \subfigure[Scenario 8]{
    \includegraphics[width=3in]{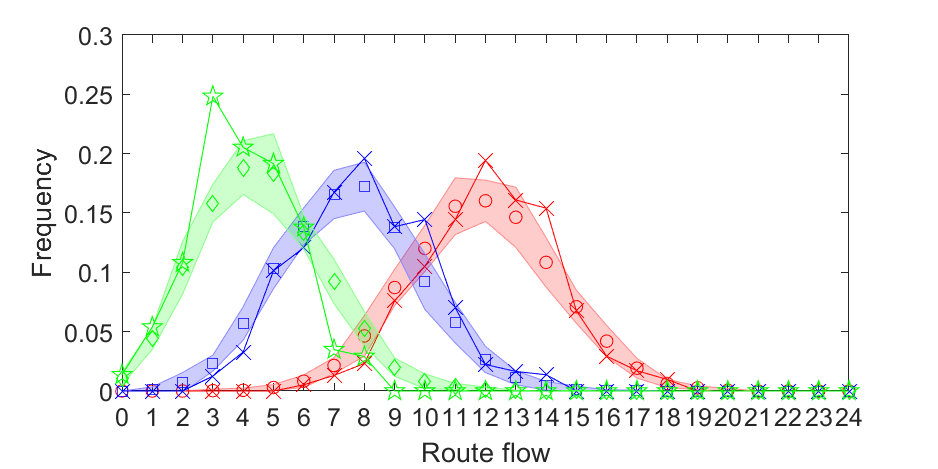}}    
    \subfigure{
    \includegraphics[width=5in]{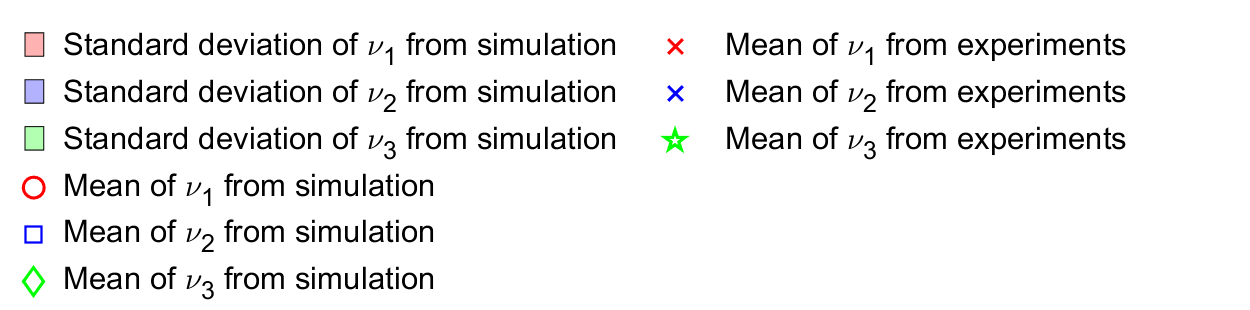}}    
    \caption{The comparison between the experimental and theoretically predicted route flow distributions.}
    \label{fig:Comparison_Distribution}
\end{figure}

Further, we compare the route flow distributions in the experiments with those deriving from the proposed model. Different from switch rates, the theoretical values of route flow distributions are difficult to be analytically calculated, since the Markov process is very complex to solve. Thus, we carry out a numerical simulation by using the calibrated parameters. 
To be consistent with the experiment as much as possible, the configuration of the simulation is set to be the same as that in the corresponding experiment, including the network configuration, the cost functions, the number of travelers, the number of rounds, and the calibrated model parameters.
The simulation is repeated 30 times for each scenario. 
A group of flow distributional data is thus obtained from every repetition.
Let $\nu_i(x)$ denote the frequency of the case that the flow of route $i$ is $x$, and the mean and standard deviation of $\nu_i$ can be calculated from the 30 groups of flow distributional data.
Results that are presented in Figure \ref{fig:Comparison_Distribution} clearly show that the proposed model well fits the experimental data, indicating that the model can describe the stochastic details of the route flow DTD evolution.

\subsection{Route-dependent Attractions}
Although the route-specific and exogenous attraction coefficients are critical to explain the experimental observations, it might bring some application concerns due to the difficulty of obtaining such a route-specific parameter in large-scale networks.
One possible solution is to link $eta_i$ with the characteristics of route $i$, which will allow us to directly calculate $eta_i$ by using the parameters of route $i$; it will avoid to arbitrarily assigning values. 
To solve this problem, more data is needed but it exceeds the scope of the paper and we leave it to future study.

Beyond that, this paper proposes the following two conjectures to explain the attraction coefficient.
(i) The attraction coefficient is determined by the travel time fluctuation of the route.
We verify the mean-variable decision rule using Table \ref{tab:Comparison}, which lists the travel time variances and the route-specific attraction coefficients in all scenarios.
It can be seen that the attraction coefficients are negatively correlated to the travel time variances in every scenario, turning out that travelers prefer low-variance routes. 
(ii) The attraction coefficient might be connected with route capacity, since travelers might prefer the routes with higher capacity.
As shown in Table \ref{tab:Comparison}, the attraction coefficients are positively correlated to the sensitivity of route cost to flow, i.e. an indicator of route capacity.

\begin{table}[!htbp]\centering\footnotesize
\setlength{\tabcolsep}{5mm}{
\caption{Relationship among marginal capacity, travel time variances, and attraction coefficient of routes.}\label{tab:Comparison}
\begin{tabular}{ccccc}
\toprule
Scenario & Route & Sensitivity of cost & Travel time variance &  Attraction coefficient $\eta_i$\\
\midrule
1 & 1 & 2 & 3.55 & 0.359 \\ 
  & 2 & 2 & 3.55 & 0.351 \vspace{2mm}\\
2  & 1 & 4 & 6.09 & 0.555 \\
   & 2 & 6 & 9.14 & 0.403 \vspace{2mm}\\
3 & 1 & 2 & 3.42 & 0.518 \\
  & 2 & 3 & 5.13 & 0.307 \vspace{2mm}\\
4 & 1 & 4 & 7.52 & 0.443 \\
  & 2 & 6 & 11.28 & 0.197 \vspace{2mm}\\
5 & 1 & 2 & 3.90 & 0.468 \\
  & 2 & 3 & 5.84 & 0.235 \vspace{2mm}\\
6 & 1 & 4 & 5.99 & 0.480 \\
  & 2 & 6 & 8.61 & 0.315 \\
  & 3 & 8 & 10.32 & 0.1250 \vspace{2mm}\\
7 & 1 & 2 & 3.34 & 0.439 \\
  & 2 & 3 & 4.64 & 0.218 \\ 
  & 3 & 4 & 5.90 & 0.0976 \vspace{2mm}\\
8 & 1 & $13.125({f_{1}}/{6}))$ & 10.11 & 0.516 \\
  & 2 & $13.125({f_{2}}/{4}))$ & 14.23 & 0.319 \\ 
  & 3 & $13.125({f_{3}}/{2}))$ & 22.93 & 0.116 \vspace{2mm}\\
\bottomrule
\end{tabular}}
\end{table}

Although the theoretical results have achieved good consistency with the experimental observations by introducing the route-dependent inertia and preference, it doesn't mean that the proposed model is the only model that could explain the experimental observations. 
Many choice behavior models exist to date, such as those in \cite{de2011expected,avineri2015prospect}, and the prospect theory that is a widely-accepted explanation of risk-aversion.
Thus, it is interesting to analyze the experimental observations using other theoretical tools.

\subsection{The Design of Laboratory Experiments}
Another discussion is regarding the design settings of the experiments. 
There is a recognized classification method of laboratory experiments proposed by the economist, Roth E. Alvin. The objective of any experiments lies in three categories, i.e., `\textit{Speaking to Theorists}' (testing and modifying formal economic theories ), `\textit{Searching for Facts}' (detecting unanticipated regularities), and `\textit{Whispering into the Ears of Princes}' (having a direct input into the policy-making process)  \citep{roth_1986}. Our experiments were intentionally designed to firstly search for behavioral regularities, and secondly test and improve existing route-choice theories accordingly.
For laboratory experiment methods, one of their important advantages is that key variables (e.g., costs, benefits, and information) could be controlled to collect 'clear' data, which is often impossible for field studies (Rapoport and Mak, 2019).
To make the best use of the advantages of experimentation, we unavoidably have to simplify  some properties of the traffic reality, e.g., network topology, cost functions, information provision mechanism, decision-making environment, etc.

%
%


\section{Conclusion}\label{sec:Conclusion}

To understand travelers' DTD route choice dynamics, we invited 312 volunteers into laboratories and conducted human-in-the-loop experiments with a variety of route choice scenarios.
Several experiment-based DTD behavioral patterns were found to be inconsistent with the existing models.
A behavioral assumption of route-dependent attractions is introduced and a discrete-choice-based analyzable DTD network flow dynamic model is proposed. 
By using the data of the experiment conducted in this research and the data reported in other studies, it is explicitly demonstrated that the proposed model is able to satisfactorily reproduce the experimental DTD dynamics in terms of the mean switching rates and the equilibrium flow of the network.
The paper is beneficial to the connection between the theoretical route choice models and human traveler's DTD route choice behaviors,
and it might be a bridge to develop a flow evolution model for real-world road networks.

Future directions may include but not limited to the following aspects.
First, more parameters and larger networks should be explored in laboratory experiments. 
However, the difficulty of calibrating parameters in route-rich networks should be carefully solved.
Second, the DTD network flow dynamic model we proposed is supposed to be examined in a wider range of datasets to test its external validity. Third, more experiments should be meticulously designed to distinguish the predictions of multiple alternative psychological mechanisms and to confirm which ones best describe the phenomenon of route-dependent attraction.

The proposed model is a deterministic model that converges to a stable and fixed value of equilibrium route flow.
Apparently, real-world network flow evolution (also exhibited by the experimental data) is oscillatory, which is difficult to describe by using a deterministic model \citep{Watling1999, He2012b, Cantarella2016,He2016b, Kumar2015, Xiao2016, Ye2017, xiao2019day-to-day}.
To overcome the limitation, the stochastic modeling method presented in Section 7.1 might be an efficient approach that is worthy to explore in the future.

\section*{Acknowledgement}

The research is funded by National Natural Science Foundation of China (72101085, 71871010) and Laboratory of Computation and Analytics of Complex Management Systems(CACMS) (Tianjin University).


\bibliographystyle{model2-names}
\bibliography{Z.Paper-D2D}

\end{spacing}
\end{document}